\documentclass[twocolumn,twoside]{IEEEtran}

\ifCLASSINFOpdf
  
\else
 
\fi

\ifCLASSOPTIONcompsoc
\usepackage[caption=false, font=normalsize, labelfont=sf, textfont=sf]{subfig}
\else
\usepackage[caption=false, font=normalsize]{subfig}
\fi
\usepackage{lipsum}%
\usepackage[dvipsnames]{xcolor}

\usepackage{balance}
\usepackage{multicol}   % for equalize of last page columns
\usepackage{cite}
\usepackage{gensymb}
\usepackage{multirow}
\usepackage{graphics}
\usepackage{epsfig}
\usepackage{graphicx}
\usepackage{epstopdf}
\usepackage{textcomp}
\usepackage{amsmath}
\usepackage{mathtools}
\usepackage{filecontents}
\usepackage{lipsum,color}
\usepackage{amssymb}
\usepackage{float}
\usepackage{times} % assumes new font selection scheme installed
\usepackage{amsthm}  % for proof 
\usepackage{amsfonts}

\theoremstyle{break}
\begin{document}
%\title{Optical Channel Modeling Between UAVs over Weak to Strong Turbulence Channels}
\title{ Channel Modeling for UAV-based Optical Wireless Links with Nonzero Boresight Pointing Errors}

%\author{Author A and Author B
\author{Mohammad~Taghi~Dabiri,~Mohsen~Rezaee,~and~Imran~Shafique~Ansari 
}    

%\markboth{IEEE~Transactions~on~Vehicular~Technology,~2020,~~~Revision}
%{       Channel Modeling for UAV-based Optical Wireless Links with Nonzero Boresight Pointing Errors}

\maketitle

%%%%%%%%%%%%%%%%%%%%%%%%%%%%%%%%%%%%%%%%%%%%%%%%%%%%%%%%%%
%%%%%%%%%%%%%%%%%%%%%%%%%%%%%%%%%%%%%%%%%%%%%%%%%%%%%%%%%%
\begin{abstract}
%%%%%%%%%%%%%%%%%%%%%%%%%%%%%%%%%%%%%%%%%%%%%%%%%%%%%%%%%%
%%%%%%%%%%%%%%%%%%%%%%%%%%%%%%%%%%%%%%%%%%%%%%%%%%%%%%%%%%
The channel modeling of unnamed aerial vehicle (UAV)-based free-space optical (FSO) links with nonzero boresight pointing error is the subject of this paper.
%We consider general case wherein the variances of UAV orientation fluctuations in $x-z$
%
In particular, utilizing log-normal turbulence model, we propose a novel closed-form statistical channel model for  UAV-based FSO links that takes into account the effect of nonzero boresight pointing errors. Subsequently, utilizing Gamma-Gamma turbulence model, we propose a novel channel characterization for such links that is valid under moderate to strong turbulence conditions. 
The accuracy of the proposed models is verified via Monte-Carlo simulations.
The proposed models are more tractable and suitable for analysis of such UAV-based FSO links. 
\end{abstract}
\IEEEpeerreviewmaketitle
%%%%%%%%%%%%%%%%%%%%%%%%%%%%%%%%%%%%%%%%%%%%%%%%%%%%%%%%%%%%
%%%%%%%%%%%%%%%%%%%%%%%%%%%%%%%%%%%%%%%%%%%%%%%%%%%%%%%%%%%%
\section{Introduction}
%%%%%%%%%%%%%%%%%%%%%%%%%%%%%%%%%%%%%%%%%%%%%%%%%%%%%%%%%%%%
%%%%%%%%%%%%%%%%%%%%%%%%%%%%%%%%%%%%%%%%%%%%%%%%%%%%%%%%%%%%
%%%%%%%%%%%%%%%%%%%%%%%%%%%%%%%%
%\subsection{Background}
%%%%%%%%%%%%%%%%%%%%%%%%%%%%%%%%

\IEEEPARstart{T}{he} realization of reliable free-space optical (FSO) backhaul and fronthaul communication links between unmanned aerial vehicles (UAVs) acting as aerial base stations is a milestone for the future development of communication networks \cite{alzenad2018fso,fawaz2018uav,yan2019comprehensive}. 
Although channel modeling in the context of terrestrial FSO communications  has been studied in  \cite{farid2007outage,jurado2011unifying,garrido2015novel,yang2014free}, these studies  cannot be directly used for UAV-based FSO systems.
%------------------------
There are several works in the literature of the long-range optical communications in space and stratosphere \cite{mai2019beam,mai2018adaptive}. However, there is one main difference between long-range optical communication and short-range multi rotor UAV-based optical communications.
In all of the proposed works in the context of optical space communications, the link length is assumed to be in the order of several hundred to several thousand kilometers, and thus, the standard deviation of alignment error must be in the order of $\micro$rad. 
For instance, in \cite{mai2019beam}, the value of standard deviation of  misalignment is 100 \micro\,rad for a 100 km stratosphere link length and in \cite{mai2018adaptive}, the value of standard deviation of  misalignment is 0.05 \micro\,rad for a 200 km stratosphere link length.
As a result, for establishing such communication links, we have to use fast and precise stabilizers which are bulky and very expensive. However, due to the payload and power consumption limitations of lightweight quadcopters drones, reaching such alignment accuracy in the order of $\micro$rad may not be always possible. From the literature of short range optical communications (the link length is mainly in the order of a few hundred meters), the standard deviation of AoA fluctuations due to orientation fluctuations of lightweight multi-rotor drones is in the order of several mrad which is approximately 250-1000 times larger than the standard deviation of AoA for the space as well as the ground FOS links \cite{safi2019spatial,dabiri2019optimal}.
%----------------------
To assess the benefit of short-range multi rotor UAV-based optical communications, one important aspect is to accurately model the channel, which has been the subject of a few recent works \cite{kaadan2014multielement, kaadan2016spherical, najafi2018statistical,najafi2019statistical,dabiri2018channel,dabiri2019tractable}.
In \cite{kaadan2014multielement,kaadan2016spherical}, a novel model was presented for FSO link between two hovering UAVs with multi-element optical transceiver arrays. 
In \cite{najafi2018statistical,najafi2019statistical}, the authors derived a statistical model for UAV-based FSO system by taking into account the non-orthogonality of the laser beam and the random fluctuations of UAVs.
Considering the joint effects of UAVs' fluctuations as well as atmospheric turbulence, a novel channel model was proposed in \cite{dabiri2018channel,dabiri2019tractable} that is suitable for hovering UAV-based FSO links with zero boresight angle. However, none of the proir studies addresses the effects of nonzero boresight UAV's angle.
Moreover, the results of \cite{kaadan2014multielement, kaadan2016spherical, najafi2018statistical,najafi2019statistical,dabiri2018channel,dabiri2019tractable} are obtained for a special case wherein the UAVs have equal variances of orientation fluctuations in $x-z$ and $y-z$ axes.

%The boresight is the fixed misalignment from transmission axis.% between transmitter (Tx) and receiver (Rx).
%
%In practical situations, in addition to the atmospheric turbulence, UAVs' position and orientation fluctuations,  position estimation errors, mechanical noise, and gimbal friction lead to a fixed displacement in UAVs' orientation which is called boresight.
In addition to the UAVs' position and orientation fluctuations, in practical situations, inevitable errors such as position estimation errors and mechanical noise lead to a fixed misalignment  between transmitter (Tx) and receiver (Rx)  mounted on UAVs, which is known as boresight.
More importantly, the variances of orientation fluctuations are not equal in $x-z$ and $y-z$ axes.
%
%
%To fill this gap, in this paper we consider a general case wherein the variances of UAV oientation fluctuations in $x-z$ and $y-z$ axes can take any values. 
%Moreover, the results of this paper will demonstrate that the boresight changes the statistical channel model of UAV-based FSO links.
%
%Hence, to asses the advantages of UAV-based FSO systems, it is important to have a comprehensive and accurate channel model with nonzero boresight which is the scope of this paper.
Hence, to assess the advantages of UAV-based FSO systems, in this paper, we consider a general case wherein the variances of UAV oientation fluctuations in $x-z$ and $y-z$ axes can take any different values and we propose a comprehensive and accurate channel model by taking into account the effects of nonzero boresight.
In particular, under the weak turbulence conditions, we propose a novel and tractable channel model for the considered UAV-based FSO link over log-normal atmospheric turbulence environment that takes into account the effects of nonzero boresight as well as  UAVs' orientation and position fluctuations, atmospheric turbulence strength, optical beamwidth, link length, lens radius size, receiver's field-of-view (FOV), etc. For moderate to strong turbulence conditions, a novel closed-form statistical channel model is derived under the influence of Gamma-Gamma (GG) turbulence channel. 
%-----------
The accuracy of the proposed models is verified by performing Monte-Carlo simulations.
The developed results can therefore be applied as a benchmark for determining the optimal tunable parameters of UAV-based FSO links under different channel conditions and varying levels of UAV instability without resorting to time-consuming Monte-Carlo simulations.

%From the results of \cite{}, the standard deviation of ground links is in the order of 100 $\micro$rad and accordingly, the optimal value for Rx FOV is lower than one mrad. Similarly, for the long-range optical stratosphere and space communications, standard deviations of alignment errors must be in the order of $\micro$rad. Such accuracy in alignment of space communications is reached by exploiting the fast and precise stabilizers which are very expensive and require more power consumptions and thus, it can not be applicable in many aerial communications such as for lightweight UAVs. Accordingly, higher orientation and position fluctuations of lightweight UAVs are one of the main differences with the optical links for ground counterpart. For instance, the orientation fluctuations of 

% https://www.space-of-innovation.com/vialight-communications-receives-spin-off-award-for-successful-technology-transfer-of-dlr/

% https://www.space-of-innovation.com/vialight-communications-receives-spin-off-award-for-successful-technology-transfer-of-dlr/
%\newpage
%%%%%%%%%%%%%%%%%%%%%%%%%%%%%%%%%%%%%%%%%%%%%%%%%%%%%%%%%%%%
%%%%%%%%%%%%%%%%%%%%%%%%%%%%%%%%%%%%%%%%%%%%%%%%%%%%%%%%%%%%
\section{System Model and Main Assumptions}
%%%%%%%%%%%%%%%%%%%%%%%%%%%%%%%%%%%%%%%%%%%%%%%%%%%%%%%%%%%%
%%%%%%%%%%%%%%%%%%%%%%%%%%%%%%%%%%%%%%%%%%%%%%%%%%%%%%%%%%%%
Similar to \cite{dabiri2018channel,dabiri2019tractable}, we consider a UAV-based FSO system wherein a hovering UAV Tx (located at $[0,0,0]$ in Cartesian coordinate system $[x,y,z]$) transmits optical signals towards a hovering UAV Rx (located at $[0,0,Z]$).
By knowing the mean positions of Tx and Rx, the UAVs try to align Tx and Rx. However, due to the inherent position and orientation fluctuations of UAVs, the instantaneous position and orientation of aerial nodes are deviated from their means.
Let the random variables (RVs) $\theta_\textrm{tx}$ and $\theta_\textrm{ty}$ denote the orientation fluctuations of Tx in $x-z$ and $y-z$ planes, respectively, the RVs $\theta_\textrm{rx}$ and $\theta_\textrm{ry}$ denote the orientation fluctuations of Rx in $x-z$ and $y-z$ planes, respectively, the RVs $x_{tx}$ and $y_{ty}$ denote the position vibrations of Tx in $x-z$ and $y-z$ planes, respectively, and the RVs $x_{rx}$ and $y_{ry}$ denote the position vibrations of Rx in $x-z$ and $y-z$ planes, respectively.
Based on numerous random events related to hovering UAVs and from the central limit theorem, position and orientation deviations of UAVs are considered as Gaussian distributed \cite{dabiri2020analytical,kaadan2014multielement,dabiri20203d}.
%
% Hence, in addition to the channel fading, fluctuations in the orientation of the UAVs (due to the effect of wind, mechanical and control system flaws, antenna and BS payload, etc.) can lead to beam misalignment and adversely affect the link performance and channel capacity.

In practical situations, in addition to aforementioned UAV's fluctuations,  position estimation errors and mechanical noise lead to a fixed displacement in UAVs' orientation, which is termed as  boresight. Therefore, we have $\theta_\textrm{i}\sim\mathcal{N}(\theta_\textrm{i}',\sigma_{io}^2)$ for $i\in\{\textrm{tx},\textrm{ty}\}$, $\theta_\textrm{i}\sim\mathcal{N}(\theta_\textrm{i}',\sigma_{io}^2)$ for $i\in\{\textrm{rx},\textrm{ry}\}$, 
$x_i\sim\mathcal{N}(0,\sigma_{ip}^2)$ for $i\in\{\textrm{tx},\textrm{rx}\}$, and
$y_i\sim\mathcal{N}(0,\sigma_{ip}^2)$ for $i\in\{\textrm{ty},\textrm{ry}\}$. 
From these, $r_d$ is the radial distance between the received beam center and the Rx lens center where 
\begin{align}
\label{df2}
&r_d   = \\  
&\sqrt{(Z\tan(\theta_{tx})+x_{tx}+x_{rx})^2 +    (Z\tan(\theta_{ty})+x_{ty}+x_{ry})^2}. \nonumber
\end{align}

The optical channel model between UAVs can be formulated as
\begin{align}
\label{a1}
h       = h_l h_a h_{pg} h_{pa},
\end{align}
where $h_l$ is the channel loss, $h_a$ is the atmospheric turbulence, $h_{pg}$ is the geometrical loss due to the deviation between the received beam center and the receiver lens center, and $h_{pa}$ is the link loss induced by the angle-of-arrival (AoA) fluctuation.
%
%For weak to moderate atmospheric turbulence conditions, $h_a$ can be well modeled by log-normal distribution as
%\begin{align}
%\label{x5}
%f_{\rm L}(h_a)=  \frac{1}{2h_a \sigma_{L}\sqrt{2\pi}} \exp\left( -\frac{\left(\ln(h_a)-2\mu_{L}\right)^2}{8\sigma^2_{L}}\right),
%\end{align}
%where $\sigma^2_{L}$ and $\mu_{L}=-\sigma^2_{L}$  denote the variance and mean of log-irradiance, respectively, where $\sigma^2_{L}\simeq \sigma^2_{R}/4$ with $\sigma_R^2$ being the Rytov variance. 

%pointing2007

\begin{figure}
	\begin{center}
		\includegraphics[width=3.3 in ]{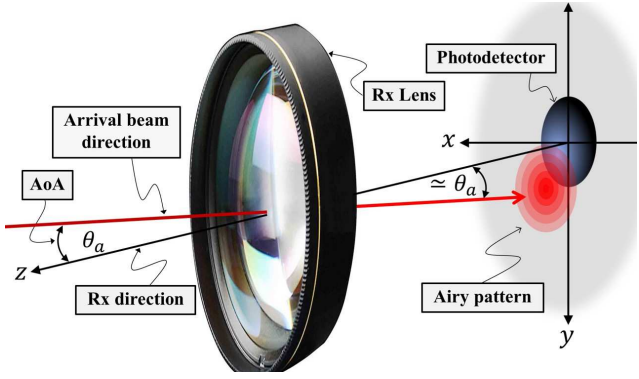}
		\caption{A schematic of the deviated optical beam due to the AoA fluctuations. The deviated  optical beam is focused by a converging lens will compose a deviated Airy pattern at the focal plane.}
		\label{sd0}
	\end{center}
\end{figure}

From \eqref{df2} and \cite[eq. (8)]{farid2007outage}, for any instantaneous value of $r_d$, the instantaneous collected optical signal by a Rx lens with radius $r_a$ (which is called geometrical pointing error coefficient) can be obtained as 
\begin{align}
\label{df3}
h_{pg} &= \int_{-r_a}^{r_a}  \int_{-\sqrt{r_a^{2}-y^2}}^{\sqrt{r_a^{2}-y^2}} \frac{2}{\pi w_{z}^2} \nonumber \\
&~~~\times e^{-2\frac{(x+\theta_{tx}Z+x_{tx}+x_{rx})^{2}+(y+\theta_{ty}Z+x_{ty}+x_{ry})^2}{w_z^2}} dx dy.
\end{align}
In addition, the AoA of the received signal is obtained as
\begin{align}
\label{df1}
\theta_a = \tan^{-1}\left(\sqrt{\left(\tan(\theta_{tx}+\theta_{rx})\right)^2+\left(\tan(\theta_{ty}+\theta_{ry})\right)^2}\right).
\end{align}
As depicted in Fig. \ref{sd0}, the collected optical signal by the converging Rx lens is guided toward a circular detector with radius $r_{ap}$. When an incident beam with small value of $\theta_a$ is passed through a lens, the outside angle of beam will be approximately unaltered \cite{gagliardi1995optical}.
As shown, a thin lens diffracts the collected light into a series of circular waves at the focal plane. The intensity of the diffracted beam pattern at the focal plane can be expressed by using the Airy pattern which is given in \cite{born2013principles}.
The fraction of collected power by the circular detector to the total power collected by the lens is the link loss induced by the AoA fluctuation which is denoted by $h_{pa}$.  From \cite{born2013principles} and \cite{mai2019beam}, for the considered system model, $h_{pa}$ can be obtained from \eqref{df4}.
%
%%%%%%%%%%%%%%%%%%%%%%%%%%%%%%%%%
%%%%%%%%%%%%%%%%%%%%%%%%%%%%%%%%%
\begin{figure*}
	\normalsize
	\begin{align}
	\label{df4}
	&h_{pa} = \frac{1}{\pi} \int_{-r_{ap}}^{r_{ap}} 
	\int_{-\sqrt{r_{ap}^2-r^2}}^{\sqrt{r_{ap}^2-x^2}} 
	\left( \frac{\mathbb{J}_1\left( \frac{\pi}{\lambda N_f} 
		\sqrt{(x-d_f\tan(\theta_{tx}+\theta_{rx}))^2+(y-d_f\tan(\theta_{ty}+\theta_{ry}))^2}\right)}
	{\sqrt{(x-d_f\tan(\theta_{tx}+\theta_{rx}))^2+(y-d_f\tan(\theta_{ty}+\theta_{ry}))^2}} \right)^2 \textrm{d}x\,\textrm{d}y.
	\end{align}
	\hrulefill
	\vspace*{4pt}
\end{figure*}
%%%%%%%%%%%%%%%%%%%%%%%%%%%%%%%%
%%%%%%%%%%%%%%%%%%%%%%%%%%%%%%%%
In \eqref{df4}, $d_f$ is the focal length,  $N_f$ is the f-number, $\lambda$ is the optical wavelength, and $\mathbb{J}_1(\cdot)$ is the Bessel function of the first kind of order one.

%
%%%%%%%%%%%%%%%%%%%%%%%%%%%%%%%%%%%%%%%%%%%%%%%%%%%%%%%%%%%%%%%%
%%%%%%%%%%%%%%%%%%%%%%%%%%%%%%%%%%%%%%%%%%%%%%%%%%%%%%%%%%%%%%%% VERSUS P_T
%%%%%%%%%%%%%%%%%%%%%%%%%%%%%%%%%%%%%%%%%%%%%%%%%%%%%%%%%%%%%%%%
%%%%%%%%%%%%%%%%%%%%%%%%%%%%%%%%%%%%%%%%%%%%%%%%%%%%%%%%%%%%%%%%
%
%

%Then, we generate 5107 independent coefficients of G(ty; ry) from an actual antenna pattern as given by (4). We also generate 5107 independent small-scale channel coefficients and calculate largescale channel coefficient from (3).

%%%%%%%%%%%%%%%%%%%%%%%%%%%%%%%%%%%%%%%%%%%%%%%
%%%%%%%%%%%%%%%%%%%%%%%%%%%%%%%%%%%%%%%%%%%%%%%
\section{Analytical Channel Modeling}
%%%%%%%%%%%%%%%%%%%%%%%%%%%%%%%%%%%%%%%%%%%%%%%
%%%%%%%%%%%%%%%%%%%%%%%%%%%%%%%%%%%%%%%%%%%%%%% 
The results of previous works in \cite{dabiri2018channel,dabiri2019tractable} are provided for a specified case wherein $\sigma_{tx}=\sigma_{ty}$, $\sigma_{rx}=\sigma_{ry}$, and $\theta_{tx}'=\theta_{ty}'=\theta_{rx}'=\theta_{ry}'=0$.
In this paper, we consider a general case with non-zero boresight angle wherein the  variances of UAV's orientation and position fluctuations are not necessarily equal in the direction of $x$ and $y$ axes, i.e., $\sigma_{tx}, \sigma_{ty}, \sigma_{rx},$ and $\sigma_{ry}$ can take any different values and $\theta_{tx}'\neq0$, $\theta_{ty}'\neq0$, $\theta_{rx}'\neq0$, $\theta_{ry}'\neq0$. 
%

%
%------------- Proposition 1 ----------------------------------------------------------------
%------------- Proposition 1 ----------------------------------------------------------------
%------------- Proposition 1 ----------------------------------------------------------------
%------------- Proposition 1 ----------------------------------------------------------------
%------------- Proposition 1 ----------------------------------------------------------------
{\bf Theorem 1.}  {\it  The distribution of link loss induced by AoA fluctuations is derived as}
%%%%%%%%%%%%%%%%%%%%%%%%%%%%%%%%%%
\begin{align}
\label{dx4}
f_{h_{pa}}(h_{pa})  = \mathbb{R}\,\delta(h_{pa}-1) + (1-\mathbb{R})\,\delta(h_{pa}),
\end{align}
{\it where $\mathbb{R}=\sum_{n=1}^{N'}\mathbb{R}_n$ and}
\begin{align}
\label{dx5}
&\mathbb{R}_n = 
\left[ Q\left(\frac{-\sqrt{\theta_\textrm{FOV}^2-(\frac{\theta_\textrm{FOV}}{N'}(n-1))^2}-\theta_{tx}'-\theta_{rx}'}{\sqrt{\sigma_{txo}^2+\sigma_{rxo}^2}}\right) \right.\nonumber \\
&~~~-
\left.Q\left(\frac{\sqrt{\theta_\textrm{FOV}^2-(\frac{\theta_\textrm{FOV}}{N'}(n-1))^2}-\theta_{tx}'-\theta_{rx}'}{\sqrt{\sigma_{txo}^2+\sigma_{rxo}^2}}\right) \right] \nonumber \\
%----------------
&~~~\times\left[
Q\left(\frac{\frac{\theta_\textrm{FOV}}{N'}(n-1)-\theta_{ty}'-\theta_{ry}'}{\sqrt{\sigma_{tyo}^2+\sigma_{ryo}^2}}\right) \right.\nonumber \\
&~~~-Q\left(\frac{\frac{n\,\theta_\textrm{FOV}}{N'}-\theta_{ty}'-\theta_{ry}'}{\sqrt{\sigma_{tyo}^2+\sigma_{ryo}^2}}\right) 
+Q\left(\frac{-\frac{n\,\theta_\textrm{FOV}}{N'}-\theta_{ty}'-\theta_{ry}'}{\sqrt{\sigma_{tyo}^2+\sigma_{ryo}^2}}\right) \nonumber \\
&~~~\left.-Q\left(\frac{-\frac{\theta_\textrm{FOV}}{N'}(n-1)-\theta_{ty}'-\theta_{ry}'}{\sqrt{\sigma_{tyo}^2+\sigma_{ryo}^2}}\right) \right] 
\end{align}
{\it where $Q(\cdot)$ and $\delta(.)$ are the well-known Q-function and Dirac delta function, respectively.}
%%%%%%%%%%%%%%%%%%%%%%%%%%% begin PROOF %%%%%%%%%%%%%%%%%%%%%%%%%%%%%%%%%%%%%%%%%%%%%%%%%%%%%%%%%%%%%%%%%%%%%%%%%%%%%%%%%%%%%%%%%%%%%%%%%%%%%%%%%%%%%%%%%%%%%%%%%%
\begin{IEEEproof}
Please refer to Appendix \ref{Apx1}.
\end{IEEEproof}
%%%%%%%%%%%%%%%%%%%%%%%%%%% END PROOF %%%%%%%%%%%%%%%%%%%%%%%%%%%%%%%%%%%%%%%%%%%%%%%%%%%%%%%%%%%%%%%%%%%%%%%%%%%%%%%%%%%%%%%%%%%%%%%%%%%%%%%%%%%%%%%%%%%%%%%%%%%%
In \eqref{dx4}, the parameter $N'$ is a positive integer and when $N'$ grows, the analytical results of \eqref{dx4} leads to the simulation results. 
%It can be observed from Fig. 4 that the bounds become tighter when N grows, as expected, and that N = 20 leads to quite tight bounds.

Next, we derive the analytical channel models for UAV-to-UAV (UU) FSO links for a wide range of weak to strong atmospheric turbulence conditions.

%---------------------------------
%---------------------------------
\subsection{For Weak to Moderate Turbulence Conditions}
%---------------------------------
%---------------------------------

%
%------------- Proposition 1 ----------------------------------------------------------------
%------------- Proposition 1 ----------------------------------------------------------------
%------------- Proposition 1 ----------------------------------------------------------------
%------------- Proposition 1 ----------------------------------------------------------------
%------------- Proposition 1 ----------------------------------------------------------------
{\bf Theorem 2.}  {\it  Under weak to moderate atmospheric turbulence conditions, a probability density function (PDF) of the considered UU channel is formulated as}
%%%%%%%%%%%%%%%%%%%%%%%%%%%%%%%%%%
\begin{align} 
\label{xc2}  
&f_h(h) =  (1-\mathbb{R})\delta(h) + \frac{c_1 w_z^2\mathbb{R}}{h \sqrt{2\pi \sigma_{L}^2}}    \\ 
& \int_0^{2\pi} \int_0^{\infty}                   x
e^{c_3\,x^2+c_2\,x}  
e^{-\frac{\left(\ln(\frac{h}{A_0h_l})+x^2-2\mu_{L}\right)^2}{8\sigma^2_{L}}}
\textrm{d}x \textrm{d}\phi, \nonumber
\end{align}
{\it where}
\begin{align}
\label{xc}
\left \{
\begin{array}{ll}
&\!\!\!\!\!\!\! c_1                = \frac{w_z^2}{8\pi\sigma_{dx}\sigma_{dy}}
\exp\left(  -\frac{Z^2\theta_{tx}'^2}{2\sigma_{dx}^2}   -\frac{Z^2\theta_{ty}'^2}{2\sigma_{dy}^2}  \right),  \\
%---------------------
&\!\!\!\!\!\!\! c_2  = \frac{w_z}{\sqrt{2}}\left(  \frac{Z\theta_{tx}' \cos(\phi)}{\sigma_{dx}^2}  +   \frac{Z\theta_{ty}' \sin(\phi)}{\sigma_{dy}^2} \right), \\
%---------------------
&\!\!\!\!\!\!\! c_3   = \frac{     w_z^2(\sigma_x^2-\sigma_y^2)\cos(2\phi) -w_z^2(\sigma_{dx}^2+\sigma_{dy}^2)     } 
{8\sigma_{dx}^2\sigma_{dy}^2}.
\end{array}
\right. ~~~~~~~~~~
\end{align}

%%%%%%%%%%%%%%%%%%%%%%%%%%% begin PROOF %%%%%%%%%%%%%%%%%%%%%%%%%%%%%%%%%%%%%%%%%%%%%%%%%%%%%%%%%%%%%%%%%%%%%%%%%%%%%%%%%%%%%%%%%%%%%%%%%%%%%%%%%%%%%%%%%%%%%%%%%%
\begin{IEEEproof}
	Please refer to Appendix \ref{Apc}.
\end{IEEEproof}
%%%%%%%%%%%%%%%%%%%%%%%%%%% END PROOF %%%%%%%%%%%%%%%%%%%%%%%%%%%%%%%%%%%%%%%%%%%%%%%%%%%%%%%%%%%%%%%%%%%%%%%%%%%%%%%%%%%%%%%%%%%%%%%%%%%%%%%%%%%%%%%%%%%%%%%%%%%%

As we will observe, the proposed channel model in \eqref{xc2} well models the optical channels between UAVs under weak to moderate turbulence condition. However, it consists of a two-dimensional integral. In the next Theorem, we try to find a more tractable channel models under weak turbulence conditions.

%------------- Proposition 1 ----------------------------------------------------------------
%------------- Proposition 1 ----------------------------------------------------------------
%------------- Proposition 1 ----------------------------------------------------------------
%------------- Proposition 1 ----------------------------------------------------------------
%------------- Proposition 1 ----------------------------------------------------------------
{\bf Theorem 3.}  {\it  Under weak to moderate atmospheric turbulence conditions, a closed-form PDF of the considered UU link is obtained as} 
\begin{align}
\label{po1}
% fh_ana_Rayl           = L3 * h.^(tau-1)  .*    qfunc( (L1*log(h/A0/h_l)-L2)  ) ;
f_h(h)       = (1-\mathbb{R})\delta(h) + v_3 \mathbb{R}h^{\tau_1-1} Q\left( v_1\ln\left(\frac{h}{A_0h_l}\right) - v_2 \right),
\end{align}
{\it where $v_1    = \frac{1}{2 \sigma_L^2}$, 
	%
	%v2   = (mu_Lnha/2/var_Lnha-tau)/v1;
$v_2= \left(\frac{\mu_L}{2 v_1 \sigma_L^2}-\frac{\tau_1}{v_1}\right)$,
	%
	%v3 = tau * (A0*h_l)^(-tau) * exp(v2^2/2 - 2*mu_Lnha^2*v1^2)   ;	
$v_3=\tau_1 (A_0h_l)^{-\tau_1}\exp\left(\frac{v_2^2}{2}-2\mu_L^2 v_1^2 \right)$, and 
$\tau_1 = \frac{w_z^2}{4\sigma_m^2}$.}

%%%%%%%%%%%%%%%%%%%%%%%%%%% begin PROOF %%%%%%%%%%%%%%%%%%%%%%%%%%%%%%%%%%%%%%%%%%%%%%%%%%%%%%%%%%%%%%%%%%%%%%%%%%%%%%%%%%%%%%%%%%%%%%%%%%%%%%%%%%%%%%%%%%%%%%%%%%
\begin{IEEEproof}
	Please refer to Appendix \ref{Ape}.
\end{IEEEproof}
%%%%%%%%%%%%%%%%%%%%%%%%%%% END PROOF %%%%%%%%%%%%%%%%%%%%%%%%%%%%%%%%%%%%%%%%%%%%%%%%%%%%%%%%%%%%%%%%%%%%%%%%%%%%%%%%%%%%%%%%%%%%%%%%%%%%%%%%%%%%%%%%%%%%%%%%%%%%

The channel model proposed in \eqref{po1} is more tractable than \eqref{xc2} and as we will observe in the next Section,  \eqref{po1} is valid over a wide range of pointing errors. However, the proposed channel model in \eqref{po1} deviates from on which obtained by simulation for a special case wherein 
$\Big[(\theta_{tx}'+\theta_{rx}')^2 + (\theta_{ty}'+\theta_{ry}')^2\Big]>\Big[9\max(\sigma_{txo}^2+\sigma_{rxo}^2,\sigma_{tyo}^2+\sigma_{ryo}^2)\Big]$.
This deviation is related to the approximation used in \eqref{kp1}.

%------------- Proposition 1 ----------------------------------------------------------------
%------------- Proposition 1 ----------------------------------------------------------------
%------------- Proposition 1 ----------------------------------------------------------------
%------------- Proposition 1 ----------------------------------------------------------------
%------------- Proposition 1 ----------------------------------------------------------------
{\bf Theorem 4.}  {\it
	%(Outage probability of  UAV-to-UAV FSO link over weak  turbulence conditions): 
	Under weak to moderate atmospheric turbulence conditions, when the UAV have proximately same $\sigma_{txo}\simeq\sigma_{tyo}=\sigma_{to}$ and $\sigma_{rxo}\simeq\sigma_{ryo}=\sigma_{ro}$, the channel PDF of the considered UU link is formulated as}
%%%%%%%%%%%%%%%%%%%%%%%%%%%%%%%%%%
\begin{align}
\label{sfg}
&f_{h}(h) = \mathbb{M}\left(\frac{\theta_d}{\sqrt{\sigma_{to}^2+\sigma_{ro}^2}},
\frac{\theta_\textrm{FOV}}{\sqrt{\sigma_{to}^2+\sigma_{ro}^2}}\right)\delta(h) \\
%-----------
&~~~+\left(1  -   \mathbb{M}\left(\frac{\theta_d}{\sqrt{\sigma_{to}^2+\sigma_{ro}^2}},
\frac{\theta_\textrm{FOV}}{\sqrt{\sigma_{to}^2+\sigma_{ro}^2}}\right)\right)\times f_h(h>0), \nonumber
\end{align}
where
\begin{align}
\label{z1}
f_{h}(h>0) =
\left \{
\begin{array}{ll}
&\!\!\!\!\!\!\! f^a_{h}(h)~~~~~\textrm{for} ~~~ h\geq e^{q_2}  \\
&\!\!\!\!\!\!\! f^b_{h}(h)~~~~~\textrm{for} ~~~ 0<h< e^{q_2} 
\end{array}
\right. , 
\end{align}
and
\begin{align}
\label{z2}
\left\{
\begin{array}{ll}
&\!\!\!\!\!\!\! f^a_{h}(h) =q_{3}   \frac{e^{ \tau\ln h} }{h}    
\sum_{k=0}^K \sum_{j=0}^k q_1  
\left({q_2  -\ln h}\right)^{k-j} \\
&\!\!\!\!\!\!\! ~~~~~~\times \Gamma\left(\frac{j+1}{2},\frac{(q_2  -\ln h)^2}{8\sigma_L^2}\right),  \\
%----------------------------------------
&\!\!\!\!\!\!\! f^b_{h}(h) = q_{3}   \frac{e^{ \tau\ln h} }{h}
\sum_{k=0}^K \sum_{j=0}^k q_1  
\left({q_2  -\ln h}\right)^{k-j} \\
&\!\!\!\!\!\!\! ~~~~~~\times\left[\Gamma\left(\frac{j+1}{2},0\right) + (-1)^j\Upsilon\left(\frac{j+1}{2},\frac{(q_2  -\ln h)^2}{8\sigma_L^2}\right)\right], 
\end{array}
\right. 
\end{align}
%-----------------------------------------
and the constant $q_1$, $q_2$, and $q_3$ are
\begin{align}
\label{z3}
\left \{
\begin{array}{ll}
&\!\!\!\!\!\!\! q_1 =  \binom{k}{j}  \frac{(r_o^2 w_z^2/8\sigma_d^4)^k    \left(8\sigma_L^2\right)^{(j+1)/2}  }{\Gamma(k+1) k!},\\
&\!\!\!\!\!\!\! q_2 = \ln {\kappa h_l} + 2\mu_{L} - 4\sigma^2_{L}\tau,  \\
&\!\!\!\!\!\!\! q_3 = \frac{\tau \exp\left(  -\tau(q_2 + 2\sigma^2_{L}\tau) -{r_o^2}/{2\sigma_d^2}   \right) }{ 2\sqrt{8\pi}\sigma_{L}}.
\end{array}
\right.  ~~~~~~~~~~~~~~~~
\end{align}

%%%%%%%%%%%%%%%%%%%%%%%%%%% begin PROOF %%%%%%%%%%%%%%%%%%%%%%%%%%%%%%%%%%%%%%%%%%%%%%%%%%%%%%%%%%%%%%%%%%%%%%%%%%%%%%%%%%%%%%%%%%%%%%%%%%%%%%%%%%%%%%%%%%%%%%%%%%
\begin{IEEEproof}
	Please refer to Appendix \ref{Apee}.
\end{IEEEproof}
%%%%%%%%%%%%%%%%%%%%%%%%%%% END PROOF %%%%%%%%%%%%%%%%%%%%%%%%%%%%%%%%%%%%%%%%%%%%%%%%%%%%%%%%%%%%%%%%%%%%%%%%%%%%%%%%%%%%%%%%%%%%%%%%%%%%%%%%%%%%%%%%%%%%%%%%%%%%

{\bf Proposition 1.}  {\it
	%(Outage probability of  UAV-to-UAV FSO link over weak  turbulence conditions): 
	When $\frac{\sigma_d}{r_o}>0.8$, \eqref{z1} can be simplified as}
\begin{align}
\label{e1}
&f_{h>0}(h) = 
\frac{2  s_0}{\sqrt{\pi}h}  \left(s_1 + \frac{s_2(q_2  -\ln h)}{\sqrt{8\sigma_L^2}} \right)       
\\
%------
&+\frac{s_0}{h}     
\exp{\left(\frac{(q_2  -\ln h)^2}{8\sigma_L^2}\right)}  
\textrm{erfc}\left(-\frac{q_2  -\ln h}{\sqrt{8\sigma_L^2}}\right) \nonumber\\ 
%------
&\times\left(2+s_2+ \frac{2s_1 (q_2  -\ln h)}{\sqrt{8\sigma_L^2}} 
+  \frac{2s_2(q_2  -\ln h)^2}{8\sigma_L^2}  \right), \nonumber
\end{align}
where erfc$(.)$ is the well-known complementary error function, 
%$\sigma_d=\sigma_{dx}=\sigma_{dy}$,  
$r_o=Z\sqrt{\theta_{tx}'^2+\theta_{ty}'^2}$, 
$\sigma_d^2 = Z^2\sigma_\textrm{to}^2+\sigma_\textrm{tp}^2 + \sigma_\textrm{rp}^2$,
$s_0 = \frac{\tau \exp\left(  -  {r_o^2}/{2\sigma_d^2}   \right) }{ 4 }
e^{ -{\left(\ln \frac{\kappa h_l}{h}+2\mu_{L}\right)^2}\big/{8\sigma^2_{L}}  }$,
%----------------------
$s_1 =  \frac{\sqrt{2}\sigma_L r_o^2 w_z^2}{4\sigma_d^4  }$, and 
$s_2 =  \frac{\sigma_L^2 r_o^4 w_z^4}{16\sigma_d^8\Gamma(3) }$.
%%%%%%%%%%%%%%%%%%%%%%%%%%% begin PROOF %%%%%%%%%%%%%%%%%%%%%%%%%%%%%%%%%%%%%%%%%%%%%%%%%%%%%%%%%%%%%%%%%%%%%%%%%%%%%%%%%%%%%%%%%%%%%%%%%%%%%%%%%%%%%%%%%%%%%%%%%%
\begin{IEEEproof}
	When $\frac{\sigma_d}{r_o}>0.8$, \eqref{x8} can be approximated as
	\begin{align}
	\label{e2}
	&f_{h'}(h') \simeq \frac{\tau e^{-r_o^2/2\sigma_d^2}}{ h' \sqrt{8\pi}\sigma_{L}}     \int_0^\infty 
	\left( 1 +   \frac{r_o^2 w_z^2 x}{8\sigma_d^4}
	+ \frac{r_o^4 w_z^4 x^2}{256\sigma_d^8}     
	\right)  \\  
	&~~~~~~~\times\exp\left( -\frac{\left(x-(\ln \frac{\kappa h_l}{h'}+2\mu_{L}) \right)^2     +    8\sigma^2_{L}\tau x}
	{8\sigma^2_{L}}\right) dx. \nonumber
	\end{align}
	Using \cite[eq. (01.03.21.0104.01)]{wolfra}, \eqref{sfg1}, \eqref{e2}, \cite[eq. (21)]{dabiri2019tractable}, and after some mathematical derivations, the closed-form expressions for $f_{h}(h>0)$ is derived in \eqref{e1}.
\end{IEEEproof}
%%%%%%%%%%%%%%%%%%%%%%%%%%% END PROOF %%%%%%%%%%%%%%%%%%%%%%%%%%%%%%%%%%%%%%%%%%%%%%%%%%%%%%%%%%%%%%%%%%%%%%%%%%%%%%%%%%%%%%%%%%%%%%%%%%%%%%%%%%%%%%%%%%%%%%%%%%%%

%---------------------------------
%---------------------------------
\subsection{For Moderate to Strong Turbulence Conditions}
%---------------------------------
%---------------------------------

%------------- Proposition 1 ----------------------------------------------------------------
%------------- Proposition 1 ----------------------------------------------------------------
%------------- Proposition 1 ----------------------------------------------------------------
%------------- Proposition 1 ----------------------------------------------------------------
%------------- Proposition 1 ----------------------------------------------------------------
{\bf Theorem 5.}  {\it  Under moderate to strong atmospheric turbulence conditions, the channel PDF of the considered UU link is formulated as}
\begin{align}
\label{cc1}   
f_h(h) &=  (1-\mathbb{R})\delta(h) + c_4
h^{\frac{\alpha+\beta}{2}-1}
\int_0^{2\pi} \int_0^{A_0}    
x^{-\frac{\alpha+\beta}{2}-c_3-1} \nonumber \\  
%----------------
%----------------
&~~~\times \mathbb{R} e^{c_2\sqrt{\ln\left(\frac{A_0}{x}\right)}}
k_{\alpha-\beta}\left(\sqrt{\frac{4\alpha\beta h}{h_lx}}\right) 
\textrm{d}x \textrm{d}\phi, 
\end{align}
{\it where $c_4 = \frac{2c_1A_0^{c_3} }{\Gamma(\alpha)\Gamma(\beta)}  
	\left(\frac{\alpha\beta}{h_l}\right)^{\frac{\alpha+\beta}{2}}$}.

%%%%%%%%%%%%%%%%%%%%%%%%%%% begin PROOF %%%%%%%%%%%%%%%%%%%%%%%%%%%%%%%%%%%%%%%%%%%%%%%%%%%%%%%%%%%%%%%%%%%%%%%%%%%%%%%%%%%%%%%%%%%%%%%%%%%%%%%%%%%%%%%%%%%%%%%%%%
\begin{IEEEproof}
	Please refer to Appendix \ref{Apr}.
\end{IEEEproof}
%%%%%%%%%%%%%%%%%%%%%%%%%%% END PROOF %%%%%%%%%%%%%%%%%%%%%%%%%%%%%%%%%%%%%%%%%%%%%%%%%%%%%%%%%%%%%%%%%%%%%%%%%%%%%%%%%%%%%%%%%%%%%%%%%%%%%%%%%%%%%%%%%%%%%%%%%%%%

As we will observe in the next Section, the proposed channel model in \eqref{cc1} well models the optical channels between UAVs under moderate to strong turbulence condition. However, it consists of a two-dimensional integral. In the next Theorem, we provide a more tractable channel models under moderate to strong turbulence conditions.

%------------- Proposition 1 ----------------------------------------------------------------
%------------- Proposition 1 ----------------------------------------------------------------
%------------- Proposition 1 ----------------------------------------------------------------
%------------- Proposition 1 ----------------------------------------------------------------
%------------- Proposition 1 ----------------------------------------------------------------
{\bf Theorem 6.}  {\it  Under moderate to strong atmospheric turbulence conditions, the closed-form PDF of the considered UU link is formulated as}
\begin{align}
\label{b1}   
f_h(h) &= (1-\mathbb{R})\delta(h) + \mathbb{R}\sum_{m=0}^M 
%K4*((A0*h_l*h_m)^K2 - h.^K2)
\bigg[ k_3 \left(  (A_0h_lh_m)^{k_1} - h^{k_1} \right) \nonumber \\
&~~~- k_4 \left(  (A_0h_lh_m)^{k_2} - h^{k_2} \right)  \bigg]h^{\tau_1-1},  
\end{align}
{\it where $0<h\leq A_0h_lh_m$, $\nu_b=\alpha-\beta$, and}
\begin{align}
\left \{
\begin{array}{ll}
%m-tau_1+b
&\!\!\!\!\!\!\! k_1 = m-\tau_1+\beta,~~~~k_2=m-\tau_1+\alpha, \\
% K1*gamma(m-nu_b+1)*factorial(m)
&\!\!\!\!\!\!\! k_3= \frac{k_5 (\alpha\beta)^{m-\frac{\nu_b}{2}} (A_0h_l)^{-k_1}}
{k_1m! \Gamma(m-\nu_b+1)}, \\
%----------------------------------
&\!\!\!\!\!\!\! k_4=\frac{k_5 (\alpha\beta)^{m+\frac{\nu_b}{2}} (A_0h_l)^{-k_2}}
{k_2m! \Gamma(m+\nu_b+1)},\\
%----------------------------------
%gamma(a)*gamma(b)*sin(pi*nu_b)*(A0*h_l)^(tau_1)
&\!\!\!\!\!\!\! k_5=\frac{\pi(\alpha\beta)^{\frac{\alpha+\beta}{2}}\tau_1}
{\Gamma(\alpha)\Gamma(\beta)\sin(\pi\nu_b)(A_0h_l)^{\tau_1}}. \\
\end{array}
\right.
\end{align}
Moreover, the parameters $M$ and $h_m$ are given in \cite[Table I]{dabiri2019tractable}.

%%%%%%%%%%%%%%%%%%%%%%%%%%% begin PROOF %%%%%%%%%%%%%%%%%%%%%%%%%%%%%%%%%%%%%%%%%%%%%%%%%%%%%%%%%%%%%%%%%%%%%%%%%%%%%%%%%%%%%%%%%%%%%%%%%%%%%%%%%%%%%%%%%%%%%%%%%%
\begin{IEEEproof}
	Please refer to Appendix \ref{Apre}.
\end{IEEEproof}
%%%%%%%%%%%%%%%%%%%%%%%%%%% END PROOF %%%%%%%%%%%%%%%%%%%%%%%%%%%%%%%%%%%%%%%%%%%%%%%%%%%%%%%%%%%%%%%%%%%%%%%%%%%%%%%%%%%%%%%%%%%%%%%%%%%%%%%%%%%%%%%%%%%%%%%%%%%%

The channel model proposed in \eqref{b1} is more tractable than \eqref{cc1} and as we will observe in the next Section,  \eqref{b1} is valid over a wide range of pointing errors. However, the proposed channel model in \eqref{b1} deviates from on which obtained by simulation for a special case wherein 
$\Big[(\theta_{tx}'+\theta_{rx}')^2 + (\theta_{ty}'+\theta_{ry}')^2\Big]>\Big[9\max(\sigma_{txo}^2+\sigma_{rxo}^2,\sigma_{tyo}^2+\sigma_{ryo}^2)\Big]$.
This deviation is related to the approximation used in \eqref{kp1}.

%As we will observe, under weak to moderate turbulence, the proposed model is accurate. 
%The proposed model is a function of severety of UAVs' orientation and position fluctuations, boresight angle, link length, optical beamwidth, receiver lens radius, receiver's FOV, atmospheric turbulence strength, channel loss, etc.  

%------------- Proposition 1 ----------------------------------------------------------------
%------------- Proposition 1 ----------------------------------------------------------------
%------------- Proposition 1 ----------------------------------------------------------------
%------------- Proposition 1 ----------------------------------------------------------------
%------------- Proposition 1 ----------------------------------------------------------------

%------------- Proposition 1 ----------------------------------------------------------------
%------------- Proposition 1 ----------------------------------------------------------------
%------------- Proposition 1 ----------------------------------------------------------------
%------------- Proposition 1 ----------------------------------------------------------------
%------------- Proposition 1 ----------------------------------------------------------------
{\bf Theorem 7.}  {\it
	%(Outage probability of  UAV-to-UAV FSO link over weak  turbulence conditions): 
	Under moderate to strong atmospheric turbulence conditions, the channel PDF of considered UU link is formulated as}
%%%%%%%%%%%%%%%%%%%%%%%%%%%%%%%%%%
\begin{align}
\label{d1}
&f_{h}(h) = \mathbb{M}\left(\frac{\theta_d}{\sqrt{\sigma_{to}^2+\sigma_{ro}^2}},
\frac{\theta_\textrm{FOV}}{\sqrt{\sigma_{to}^2+\sigma_{ro}^2}}\right)\delta(h) \\
%-----------
&~~~+\left(1  -   \mathbb{M}\left(\frac{\theta_d}{\sqrt{\sigma_{to}^2+\sigma_{ro}^2}},
\frac{\theta_\textrm{FOV}}{\sqrt{\sigma_{to}^2+\sigma_{ro}^2}}\right)\right)\times f_h(h>0), \nonumber
\end{align}
where
\begin{align}
&f_{h}(h>0)=       
\sum_{m=0}^M  \sum_{k=0}^K \sum_{j=0}^k g_0 g_{1k}    
(h)^{\gamma-1}  \left( \ln \left( \frac{\kappa h_l h_m}{h}\right) \right)^{k-j}  \nonumber\\
%------------------------------------------------------------------------------
&+     
\sum_{m=0}^M \sum_{k=0}^K g_{1k} \left(  g_{3m} g_{7k} (h)^{m+\alpha-1} - g_{2m} g_{6k} (h)^{m+\beta-1}  \right) 
\end{align}
{\it and}
\begin{align}
\left \{
\begin{array}{ll}
&\!\!\!\!\!\!\! g_0 = (g_{2m} g_{4j} -g_{3m} g_{5j}), ~~~~~~~~~ \\
&\!\!\!\!\!\!\! g_{1k} = \frac{\pi \gamma  \left({r_o^2 w_z^2}\big/{8\sigma_d^4}\right)^k  \exp(-r_o^2/2\sigma_d^2)}
{ k!\Gamma(k+1) \Gamma(\alpha)\Gamma(\beta)\sin(\pi (\alpha-\beta))}, \\
%-------------
&\!\!\!\!\!\!\! g_{2m} = \frac{(\alpha\beta/\kappa h_l)^{m+\beta} }{\Gamma(m+\beta-\alpha+1)m!},~~ 
g_{3n} = \frac{(\alpha\beta/\kappa h_l)^{m+\alpha}  }{\Gamma(m+\alpha-\beta+1)n!}, \\
%----------------------------------
&\!\!\!\!\!\!\! g_{4j} = \frac{(-1)^j j! \binom{k}{j} \left({\kappa h_l h_m}\right)^{m+\beta-\gamma} }{(m+\beta-\gamma)^{j+1}}, ~~~~~~~~~~~~~~~~~~~~~~\\
&\!\!\!\!\!\!\! g_{5j} = \frac{(-1)^j j! \binom{k}{j} \left({\kappa h_l h_m}\right)^{m+\alpha-\gamma} }{(m+\alpha-\gamma)^{j+1}}, \\
%-----------------------------------
&\!\!\!\!\!\!\! g_{6k} = \frac{(-1)^k k!}{(n+\beta-\gamma)^{k+1}}, ~~
g_{7k} = \frac{(-1)^k k!}{(n+\alpha-\gamma)^{k+1}}. \\
\end{array}
\right.
\end{align}
%%%%%%%%%%%%%%%%%%%%%%%%%%% begin PROOF %%%%%%%%%%%%%%%%%%%%%%%%%%%%%%%%%%%%%%%%%%%%%%%%%%%%%%%%%%%%%%%%%%%%%%%%%%%%%%%%%%%%%%%%%%%%%%%%%%%%%%%%%%%%%%%%%%%%%%%%%%
\begin{IEEEproof}
 Please refer to Appendix \ref{Apf}
\end{IEEEproof}
%%%%%%%%%%%%%%%%%%%%%%%%%%% END PROOF %%%%%%%%%%%%%%%%%%%%%%%%%%%%%%%%%%%%%%%%%%%%%%%%%%%%%%%%%%%%%%%%%%%%%%%%%%%%%%%%%%%%%%%%%%%%%%%%%%%%%%%%%%%%%%%%%%%%%%%%%%%%

%\textcolor{red}{the parameters $N$ and $h_m$ must be defined.}

{\it Remark 1.} The channel model provided in this paper are for the general UU link. The results can be used for the special case of the ground-to-UAV link by setting the parameters related to the Tx orientation fluctuations to zero. Similarly, the results can be used for UAV-to-ground link by setting the parameters related to the Rx orientation fluctuations to zero.

%%%%%%%%%%%%%%%%%%%%%%%%%%%%
\begin{figure}
	\centering
	\subfloat[] {\includegraphics[width=1.67 in]{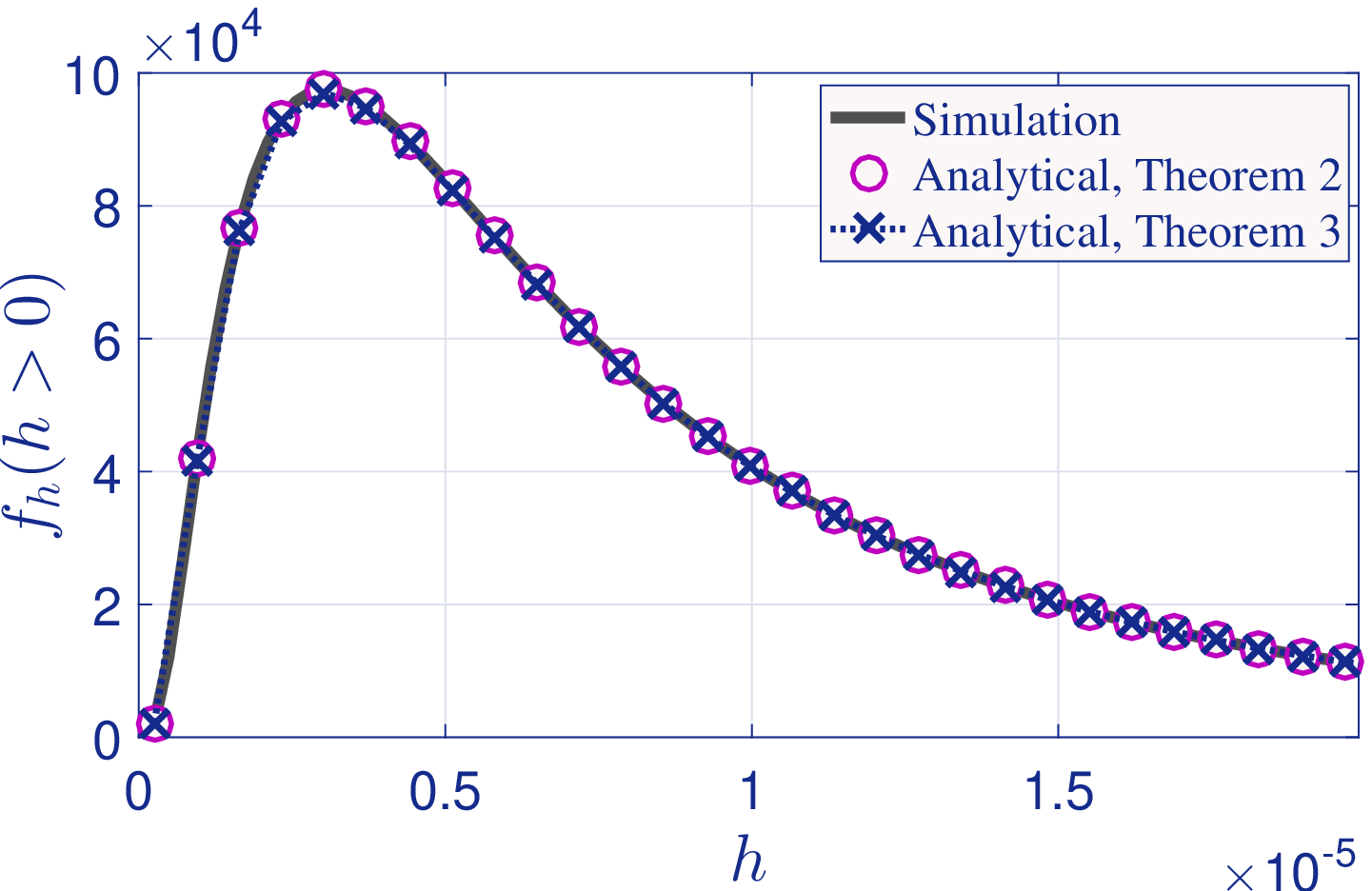}
		\label{log_23}
	}
	\hfill
	\subfloat[] {\includegraphics[width=1.67 in]{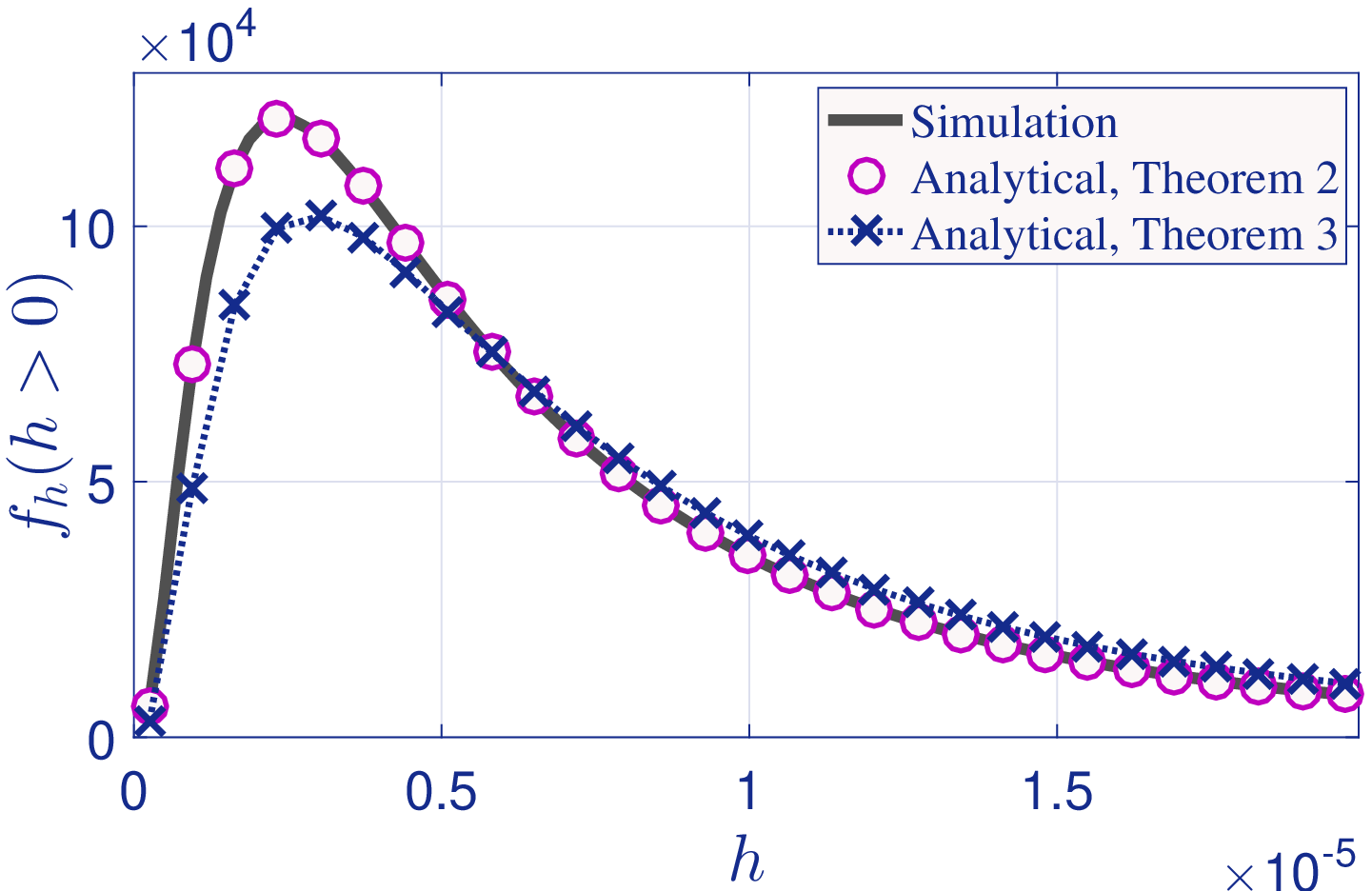}
		\label{log_24}
	}
	\caption{Comparison of the accuracy of channel PDFs given in Theorems 2 and 3  when $\sigma_{txo}=3,\sigma_{tyo}=4,\sigma_{rxo}=3$, and $\sigma_{ryo}=2$ mrad and for (a) moderate boresight with $\theta_{tx}'=2,\theta_{ty}'=3,\theta_{rx}'=2$ and $\theta_{ry}'=3$ mrad, and (b) higher boresight with $\theta_{tx}'=9,\theta_{ty}'=7,\theta_{rx}'=5$ and $\theta_{ry}'=6$ mrad.}
	\label{log_233}
\end{figure}
%%%%%%%%%%%%%%%%%%%%%%%%%%%%%%%%%%%%%%%%%%%%%%%%%%%%%%%%%%%%%%%
%
%

%%%%%%%%%%%%%%%%%%%%%%%%%%%%
\begin{figure}
	\centering
	\subfloat[] {\includegraphics[width=1.67 in]{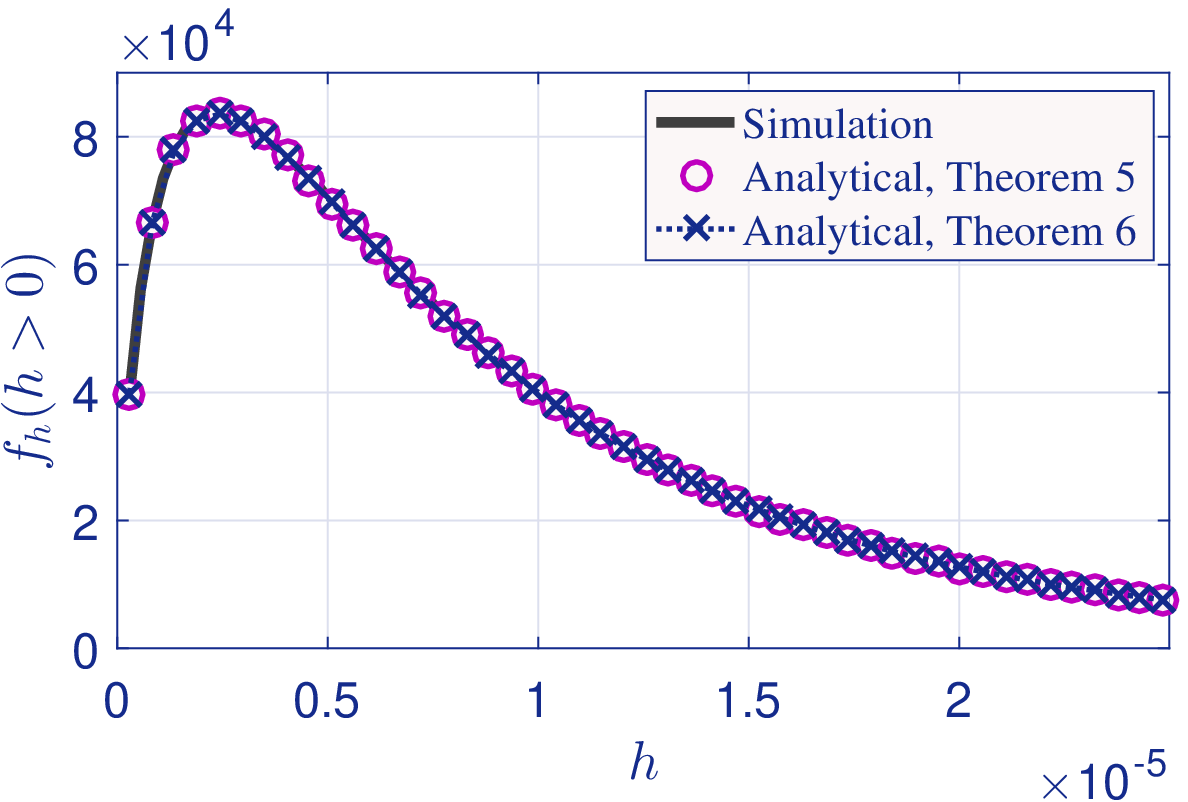}
		\label{gg_23}
	}
	\hfill
	\subfloat[] {\includegraphics[width=1.67 in]{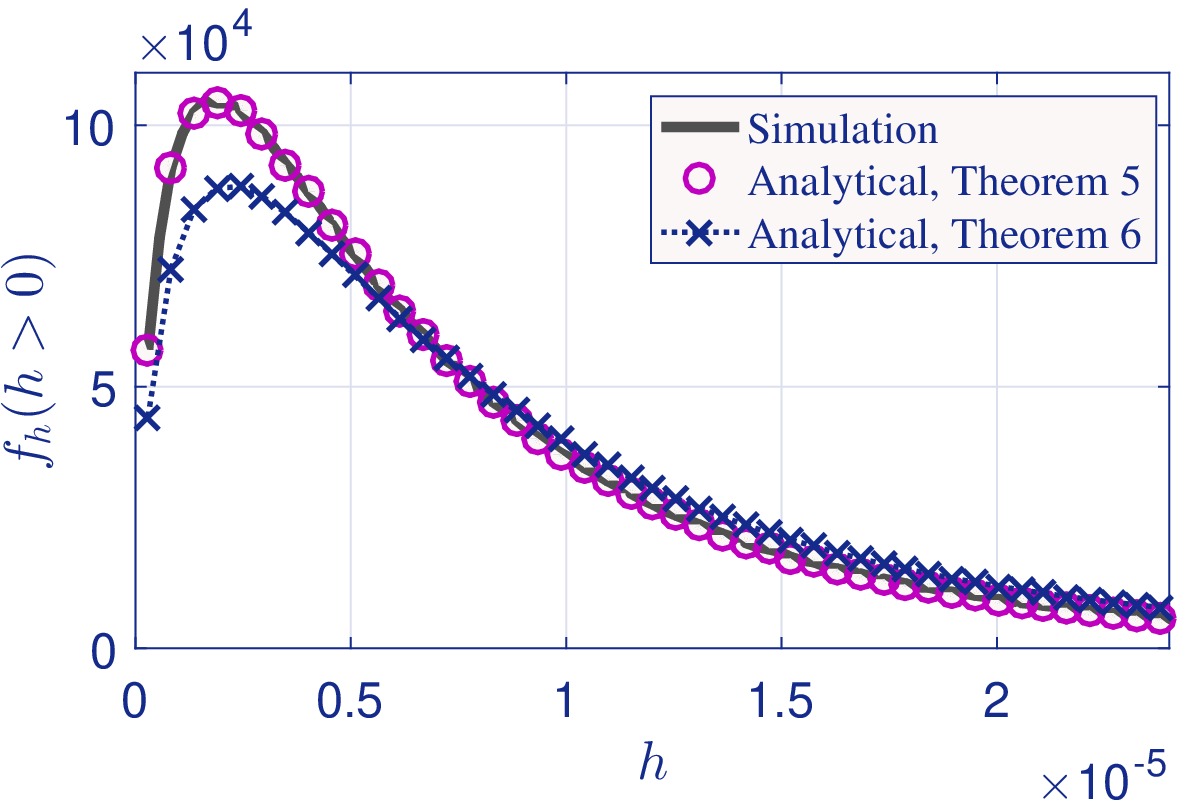}
		\label{gg_24}
	}
	\caption{Comparison of the accuracy of channel PDFs given in Theorems 5 and 6  when $\sigma_{txo}=3,\sigma_{tyo}=4,\sigma_{rxo}=3$, and $\sigma_{ryo}=2$ mrad and for (a) moderate boresight with $\theta_{tx}'=2,\theta_{ty}'=3,\theta_{rx}'=2$ and $\theta_{ry}'=3$ mrad, and (b) higher boresight with $\theta_{tx}'=9,\theta_{ty}'=7,\theta_{rx}'=5$ and $\theta_{ry}'=6$ mrad.}
	\label{gg_233}
\end{figure}
%%%%%%%%%%%%%%%%%%%%%%%%%%%%%%%%%%%%%%%%%%%%%%%%%%%%%%%%%%%%%%%

%%%%%%%%%%%%%%%%%%%%%%%%%%%%%%%%%%%%%%%%%%%%%%%%%%%%%%%%%%%%%%%%
%%%%%%%%%%%%%%%%%%%%%%%%%%%%%%%%%%%%%%%%%%%%%%%%%%%%%%%%%%%%%%%% VERSUS P_T
\begin{figure}
	\begin{center}
		\includegraphics[width=2.1 in ]{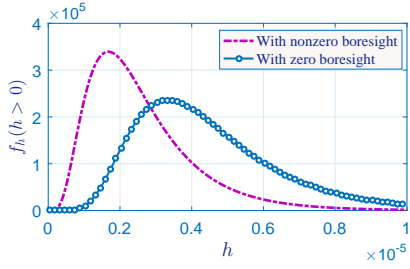}
		\caption{Comparison of channel PDF of UAV-based FSO links with zero and nonzero boresight for $\sigma_{to}=\sigma_{ro}=4$ mrad. For nonzero boresight, $\theta'_i=8$ mrad where $i\in\{\textrm{tx},\textrm{ty},\textrm{rx},\textrm{ry}\}$.}
		\label{Sn0}
	\end{center}
\end{figure}
%%%%%%%%%%%%%%%%%%%%%%%%%%%%%%%%%%%%%%%%%%%%%%%%%%%%%%%%%%%%%%%%
%%%%%%%%%%%%%%%%%%%%%%%%%%%%%%%%%%%%%%%%%%%%%%%%%%%%%%%%%%%%%%%%
%
%

%%%%%%%%%%%%%%%%%%%%%%%%%%%%
\begin{figure}
	\centering
	\subfloat[] {\includegraphics[width=1.67 in]{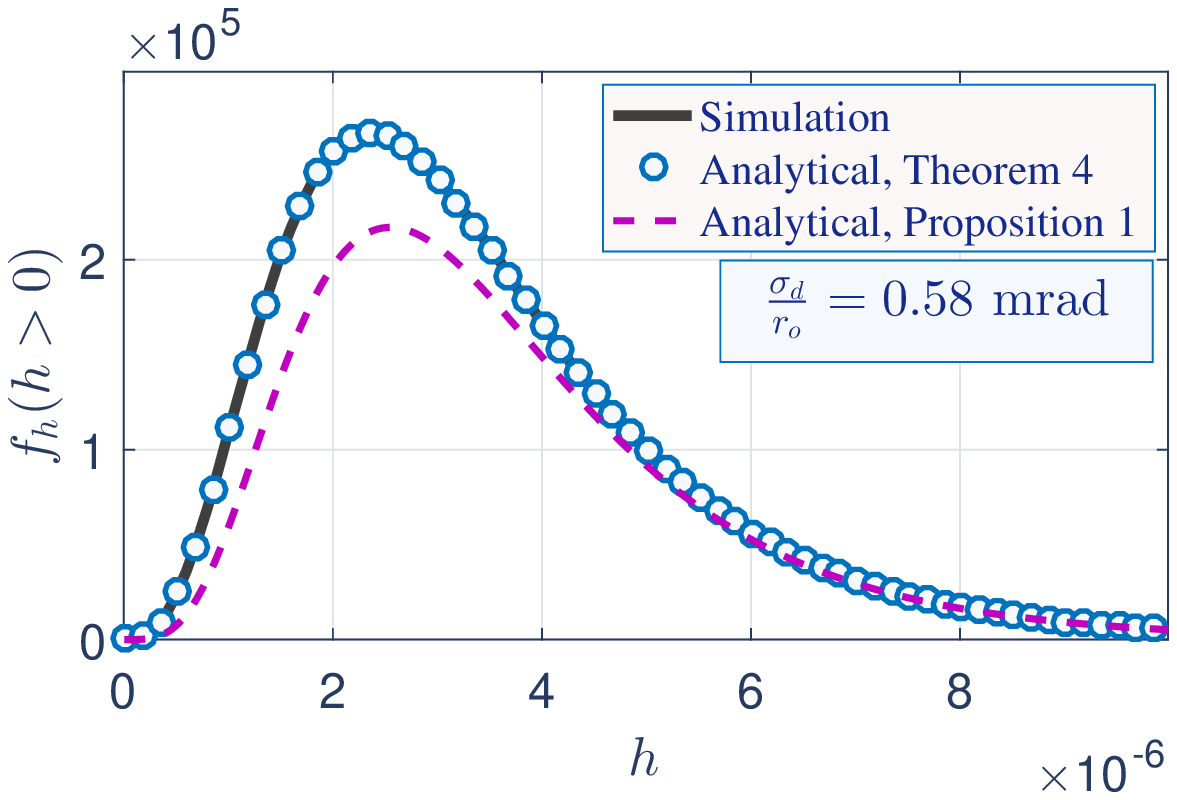}
		\label{log_1}
	}
	\hfill
	\subfloat[] {\includegraphics[width=1.67 in]{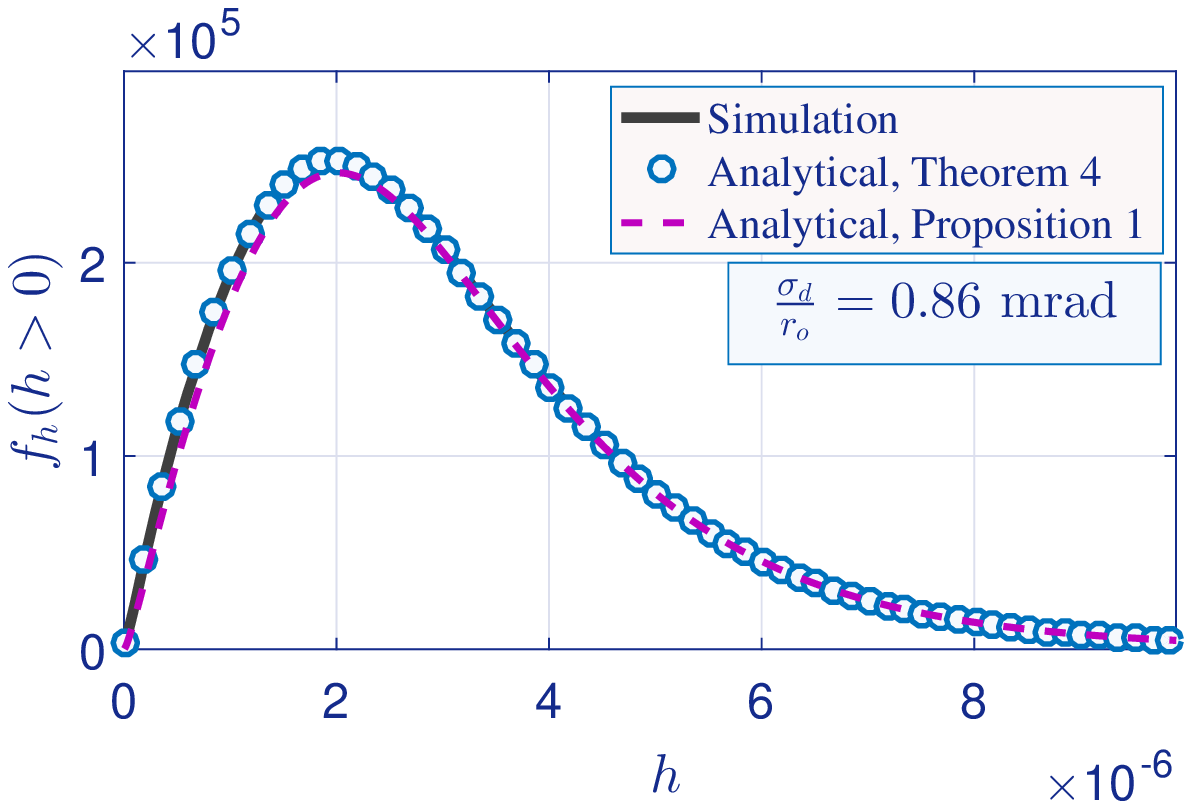}
		\label{log_2}
	}
	\caption{Comparison of the accuracy of channel PDFs given in Theorem 4 and Proposition 1  for (a) $\sigma_{to}=\sigma_{ro}=4$ mrad, and (b) $\sigma_{to}=\sigma_{ro}=6$ mrad.}
	\label{log}
\end{figure}
%%%%%%%%%%%%%%%%%%%%%%%%%%%%%%%%%%%%%%%%%%%%%%%%%%%%%%%%%%%%%%%

%-----------------------------------------
%-----------------------------------------
\section{Numerical Results}
%-----------------------------------------
%-----------------------------------------
We utilize computer simulations to verify the accuracy of our proposed analytical channel models for UAV-based FSO links.
We set the system parameters under simulation as link length $Z=500$ m, receiver lens radius $r_a =5$ cm,   Rytov variance for weak turbulence $\sigma_R^2=0.2$, for strong turbulence $\sigma_R^2=2$, standard deviation of UAV position $\sigma_{txp}=\sigma_{rxp}=40$ cm, $\sigma_{typ}=\sigma_{ryp}=30$ cm, $N'=10$, and $K=10$. Moreover, the parameters $N$ and $h_m$ are given in \cite[Table I]{dabiri2019tractable}.

For evaluation of analytical channel models provided in Section III, we perform Monte-Carlo simulations. The details of the simulation process are described as follows. 
For given $\theta_i'$ and $\sigma_{io}$ where $i\in\{tx,ty,rx,ry\}$, we generate $10^7$ independent RVs $\theta_{tx}$, $\theta_{ty}$, $\theta_{rx}$, and $\theta_{ry}$. Then, based on \eqref{df4}, we generate $10^7$ independent coefficients of $h_{pa}$.  
Moreover, for given $\sigma_{ip}$, we generate $10^7$ independent RVs $x_{tx}$, $y_{ty}$, $x_{rx}$, and $x_{ry}$. Then, using generated RVs $\theta_{tx},\theta_{ty},x_{tx},x_{ty},x_{rx}$ and $x_{ry}$, we generate $10^7$ independent coefficients of $h_{pg}$ from \eqref{df3}. For a given $\sigma_R^2<0.5$, we also generate $10^7$ independent coefficients of $h_a$ which have log-normal distribution as given in \eqref{x5}. For a given $\sigma_R^2>0.5$, we generate $10^7$ independent coefficients of $h_a$ which have GG distribution as given in \eqref{fg1}. We then obtain $10^7$ independent values of UAV-based optical channel coefficients based on \eqref{a1}. Finally, we find the channel distribution diagrams.
It is worth mentioning that, for each state of simulation, we perform independent runs in MATLAB which takes about 20 minutes of processing time (Intel Core i7 Processors, 8 GB RAM). On the other hand, by using our proposed analytical-based methods proposed in Section III, the channel can be easily modeled in less than a second which is extremely faster than employing simulation-based methods.

%log_23,log_24,log_233

First, in Fig. \ref{log_233}, we corroborate the accuracy of the derived analytical channel model in Theorems 2 and 3.
%+++The results of Fig. \ref{} are plotted for a wide range of parameter values related to orientation deviations which have greatest impact on a UAV-based optical link. 
The results of Figs. \ref{log_23} and \ref{log_24} are plotted for $\sigma_{txo}=3,\sigma_{tyo}=4,\sigma_{rxo}=3$, and $\sigma_{ryo}=2$ mrad and for a wide range of boresight values: (a) moderate boresight with $\theta_{tx}'=2,\theta_{ty}'=3,\theta_{rx}'=2$ and $\theta_{ry}'=3$ mrad, and (b) higher boresight with $\theta_{tx}'=9,\theta_{ty}'=7,\theta_{rx}'=5$ and $\theta_{ry}'=6$ mrad. The results of Figs. \ref{log_23} and \ref{log_24} clearly show that the analytical channel model derived in Theorem 2 is valid for all conditions. In Theorem 3, we also propose a more tractable closed-form channel model. As previously mentioned, the analytical channel model derived in Theorem 3 is accurate over a wide conditions, expect a specific condition wherein $\big[(\theta_{tx}'+\theta_{rx}')^2 + (\theta_{ty}'+\theta_{ry}')^2\big]>\big[9\max(\sigma_{txo}^2+\sigma_{rxo}^2,\sigma_{tyo}^2+\sigma_{ryo}^2)\big]$. The results of Fig. \ref{log_23} confirm the accuracy of expression given in Theorem 3. However, for the aforementioned specific condition, the analytical channel model given in 	Theorem 3 deviates from simulation results. 
Notice, the channel models given in Theorems 2 and 3 are provided for weak to moderate atmospheric turbulence conditions.
% %gg_23,gg_24, gg_233
Similarly, for moderate to strong turbulence conditions, in Fig. \ref{gg_233}, we corroborate the accuracy of the derived analytical channel models in Theorems 5 and 6.
The parameter values related to the UAVs' orientation fluctuations of Fig. \ref{gg_233} are equal to the parameter values used in Fig. \ref{log_233}. Simulation results confirm the accuracy of analytical channel model given in Theorem 5. Also, the closed-form channel model derived in Theorem 6 is accurate over wide range of UAVs' orientation fluctuations, expect a specific condition wherein $\big[(\theta_{tx}'+\theta_{rx}')^2 + (\theta_{ty}'+\theta_{ry}')^2\big]>\big[9\max(\sigma_{txo}^2+\sigma_{rxo}^2,\sigma_{tyo}^2+\sigma_{ryo}^2)\big]$.

%agreement with existing results in the literature and also corroborate the correctness of our proposed results.

%Following in line with objectives of the paper outlined in the introduction, we will avoid repeating the results already presented in [22] to avoid redundancy.

%
%Comparison of the accuracy of channel PDFs given in Theorems 2 and 3  when $\sigma_{txo}=3,\sigma_{tyo}=4,\sigma_{rxo}3$, and $\sigma_{ryo}=2$ mrad and for (a) moderate boresight with $\theta_{tx}'=2,\theta_{ty}'=3,\theta_{rx}'=2$ and $\theta_{ry}'=3$ mrad, and (b) higher boresight with $\theta_{tx}'=9,\theta_{ty}'=7,\theta_{rx}'=5$ and $\theta_{ry}'=6$ mrad.

In Fig. \ref{Sn0}, we compare the channel distribution of considered UAV-based system with zero and nonzero boresight pointing errors with same $\sigma_{txo}=\sigma_{tyo}=\sigma_{rxo}=\sigma_{ryo}=4$ mrad. This figure clearly shows that we can not neglect the effect of boresight pointing errors in UAV-based FSO communications even when $\sigma_{txo}=\sigma_{tyo}$ and $\sigma_{txo}=\sigma_{tyo}$,  and reveals the importance of Theorems 4 and 7. 

\begin{figure}
	\centering
	\subfloat[] {\includegraphics[width=1.67 in]{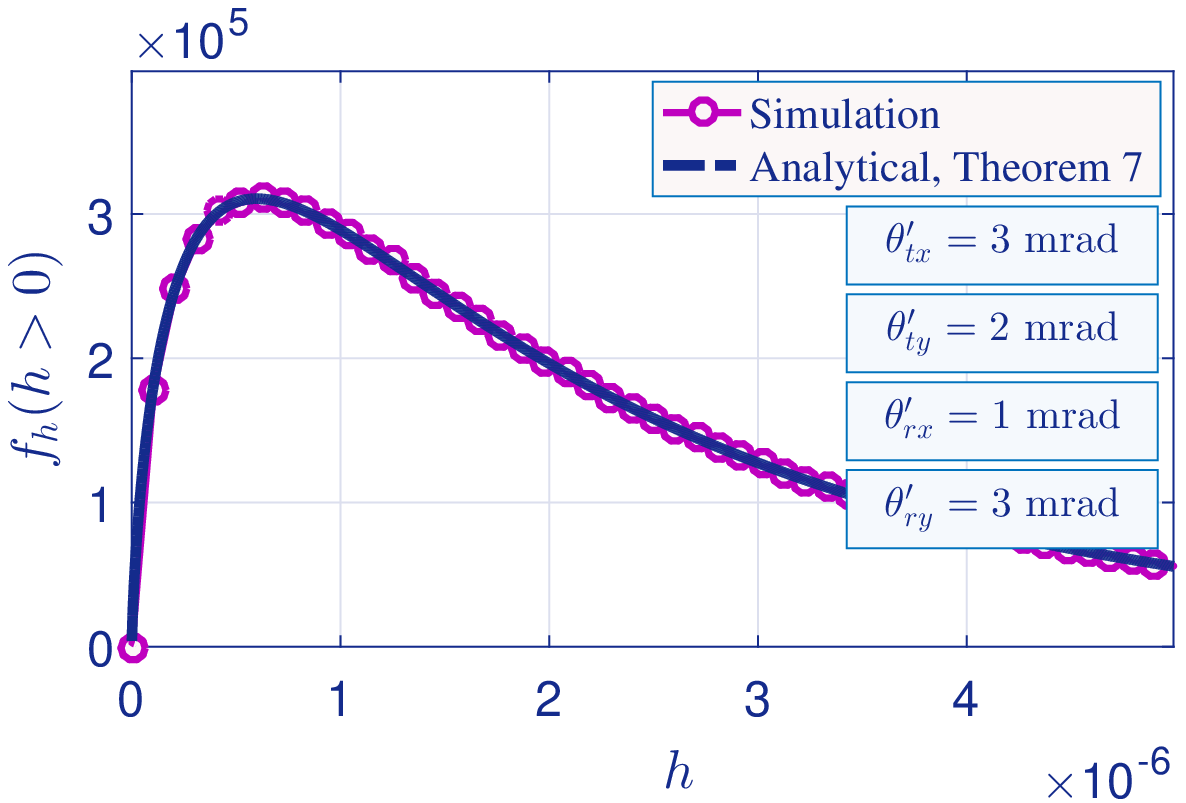}
		\label{gg_1}
	}
	\hfill
	\subfloat[] {\includegraphics[width=1.67 in]{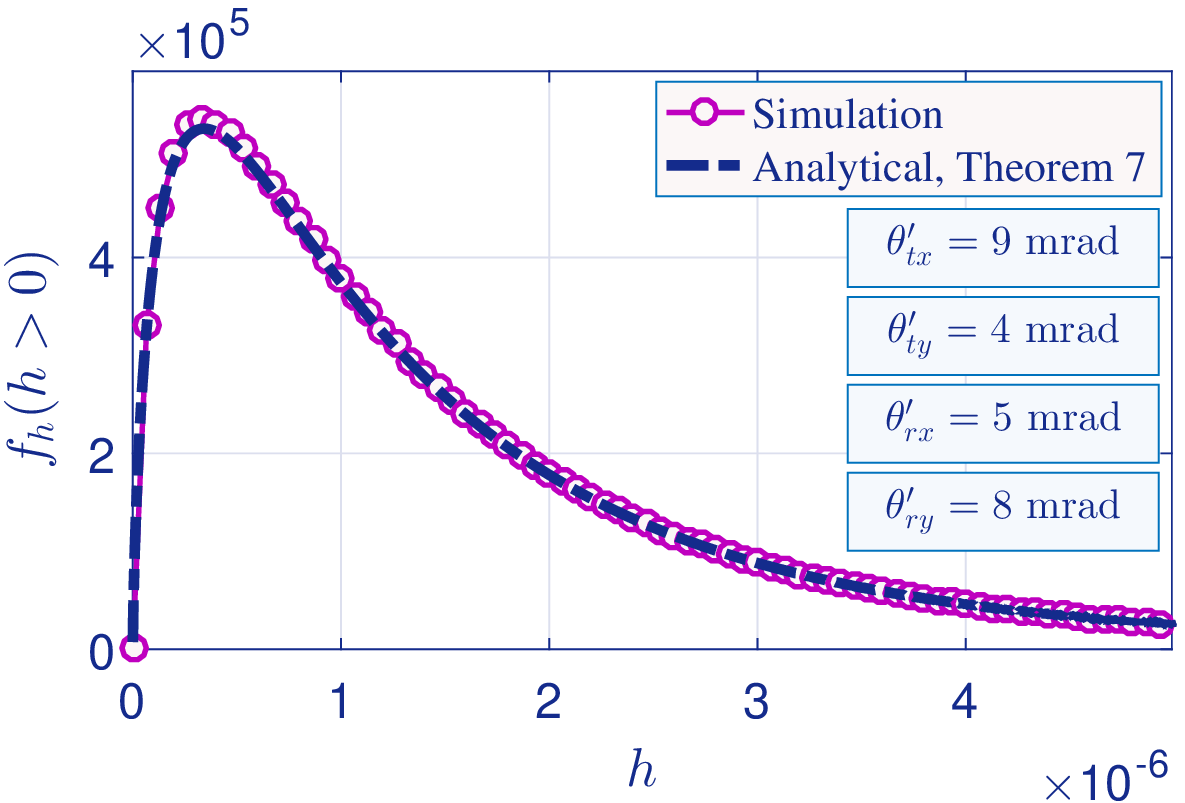}
		\label{gg_2}
	}
	\caption{Comparison of the accuracy of channel PDFs given in Theorem 7  when $\sigma_{to}=\sigma_{ro}=5$ mrad and for (a) low boresight, and (b) higher boresight.}
	\label{gg}
\end{figure}

In Fig. \ref{log}, by employing Monte-Carlo simulations, the accuracy of proposed closed-form channel PDFs under weak to moderate turbulence conditions given in Theorem 4 and Proposition 1, is investigated for two different conditions. The results are obtained for angular boresight $\theta'_i=5$ mrad  where $i\in\{\textrm{tx},\textrm{ty},\textrm{rx},\textrm{ry}\}$. The results of Fig. \ref{log} confirm the validity of analytical channel PDF proposed in Theorem 4. A more simpler channel PDF is also proposed in Proposition 1. As proven and demonstrated in Fig. \ref{log}, the simpler channel model is valid when $\frac{\sigma_d}{r_d}>0.8$. 
In Theorem 7, we proposed a channel model that is suitable for moderate to strong turbulence conditions. The results of Fig. \ref{gg} confirm the accuracy of the proposed channel PDF. 
%-----------------------------------------
%-----------------------------------------
\section{Conclusion}
%-----------------------------------------
%-----------------------------------------
In this paper, we proposed comprehensive and novel channel models for UAV-based FSO links that takes into account the effects of nonzero boresight pointing errors along with the effects of UAVs' orientation and position fluctuations, atmospheric turbulence strength, optical beamwidth, link length, lens radius size, receiver's FOV, etc. 
In addition to the tractability, simulation results confirm the accuracy of the proposed analytical channel models. 
To assess the benefits of UAV-based FSO deployments, the proposed channel models will assist researchers to easily analyze and design of such systems without using any time-consuming simulations.

%%%%%%%%%%%%%%%%%%%%%%%%%%%%
%%%%%%%%%%%%%%%%%%%%%%%%%%%%%%%%%%%%%%%%%%%%%%%%%%%%%%%%%%%%%%%

\appendices

%%-------------------------------------------------
%%-------------------------------------------------
\section{Proof of Theorem 1}
\label{Apx1}
%%-------------------------------------------------
%%-------------------------------------------------
Since the AoA angle is in the order of mrad, we can well approximate \eqref{df1} as
\begin{align}
\label{dx1}
\theta_a \simeq \sqrt{(\theta_{tx}+\theta_{rx})^2+(\theta_{ty}+\theta_{ry})^2}.
\end{align}
We consider a nonzero boresight error for AoA, and model $\theta_{tx}, \theta_{ty}, \theta_{rx},$ and $\theta_{ry}$ as nonzero mean Gaussian distributed RVs. Hence, from \eqref{dx1}, the angle $\theta_a$ follows the Beckmann distribution \cite{beckmann1987scattering}
\begin{align}
\label{dx2}
&f_{\theta_a}(\theta_a) = \frac{\theta_a}{2\pi \sqrt{(\sigma_{txo}^2+\sigma_{rxo}^2)(\sigma_{tyo}^2+\sigma_{ryo}^2)}} 
\\
&~~~~~~\times\int_0^{2\pi}  e^{- \frac{(\theta_a\cos(\phi)-\theta_{tx}'-\theta_{rx}')^2}{2(\sigma_{txo}^2+\sigma_{rxo}^2)} 
	-\frac{(\theta_a\sin(\phi)-\theta_{ty}'-\theta_{ry}')^2}{2(\sigma_{tyo}^2+\sigma_{ryo}^2)}
}\textrm{d}{\phi}. \nonumber
\end{align}
%rx and ry as nonzero mean Gaussian distributed RVs, i.e., rx ∼ N(μx, σ2x ), ry ∼ N(μy, σ2 y). Then the radial displacement r = |r| =  r2x + r2 y follows the Beckmann distribution [26]
%
As discussed in Section II, the AoA fluctuations of the hovering lightweight UAVs is in the order of several mrad which is much greater than the optical ground links. To compensate the greater AoA fluctuations, the detector area of the Rx must be selected greater than the detector area of the ground optical links, which makes a greater FoV. From, the results of \cite{dabiri2019tractable}, for a large value of FOV, one can approximate \eqref{df4} as
\begin{align}
\label{x1}
h_{pa} = 
\left \{
\begin{array}{ll}
&\!\!\!\!\!\!\! 1~~~~~~~\textrm{for}~~\theta_a<\theta_\textrm{FOV}\\
&\!\!\!\!\!\!\! 0~~~~~~~\textrm{for}~~\theta_a\geq\theta_\textrm{FOV}.
\end{array} \right.
\end{align}
where $\theta_\textrm{FOV}$ is the receiver's FOV. Now, from \eqref{dx2} and \eqref{dx1}, we have
\begin{align}
\label{dx3}
&f_{h_{pa}}(h_{pa}) = \frac{\delta(h_{pa}-1)}{2\pi \sqrt{(\sigma_{txo}^2+\sigma_{rxo}^2)(\sigma_{tyo}^2+\sigma_{ryo}^2)}} \times
\\
& \int_0^{\theta_\textrm{FOV}}\int_0^{2\pi}  \!\!  \theta_a\,
e^{- \frac{(\theta_a\cos(\phi)-\theta_{tx}'-\theta_{rx}')^2}{2(\sigma_{txo}^2+\sigma_{rxo}^2)} 
	-\frac{(\theta_a\sin(\phi)-\theta_{ty}'-\theta_{ry}')^2}{2(\sigma_{tyo}^2+\sigma_{ryo}^2)}}
\textrm{d}{\phi}\textrm{d}\theta_a \nonumber \\
%----------------------
%----------------------
&+\delta(h_{pa})\Bigg[ 1 -  \frac{1}{2\pi \sqrt{(\sigma_{txo}^2+\sigma_{rxo}^2)(\sigma_{tyo}^2+\sigma_{ryo}^2)}}\times \nonumber
\\
& \int_{\theta_\textrm{FOV}}^\infty \int_0^{2\pi}    \theta_a\,
e^{- \frac{(\theta_a\cos(\phi)-\theta_{tx}'-\theta_{rx}')^2}{2(\sigma_{txo}^2+\sigma_{rxo}^2)} 
	-\frac{(\theta_a\sin(\phi)-\theta_{ty}'-\theta_{ry}')^2}{2(\sigma_{tyo}^2+\sigma_{ryo}^2)}}
\textrm{d}{\phi}\textrm{d}\theta_a \Bigg].\nonumber
\end{align}
Similar to the method exploited in \cite{zhu2017distribution} and after some manipulations, $f_{h_{pa}}(h_{pa})$ is derived in \eqref{dx4}.

%------------------------------------
%------------------------------------
\section{Prof of Theorem 2}
\label{Apc}
%------------------------------------
%------------------------------------
For weak to moderate atmospheric turbulence conditions, $h_a$ can be well modeled by log-normal distribution as
\begin{align}
\label{x5}
f_{\rm L}(h_a)=  \frac{1}{2h_a \sigma_{L}\sqrt{2\pi}} \exp\left( -\frac{\left(\ln(h_a)-2\mu_{L}\right)^2}{8\sigma^2_{L}}\right),
\end{align}
where $\sigma^2_{L}$ and $\mu_{L}=-\sigma^2_{L}$  denote the variance and mean of log-irradiance, respectively, where $\sigma^2_{L}\simeq \sigma^2_{R}/4$ with $\sigma_R^2$ being the Rytov variance. 
From the results of \cite{dabiri2018channel}, to reduce the effects of Tx's orientation fluctuations, the divergence angle must be selected larger than the one used in conventional terrestrial FSO communications.  According to this, for UAV-based FSO communications, \eqref{df3} can be well approximated as
\begin{align}
\label{x3}
&h_{pg} \simeq \frac{2 r_a^2}{w_z^2} \\
%---------------------------
&~~~\times 
\exp\left(-2\frac{(Z\theta_{tx}+x_t+x_r)^2+(Z\theta_{ty}+y_t+y_r)^2}{w_z^2}\right)\nonumber
\end{align} 
where $r_a$ is radius of receiver lens, $w_z= w_0 \sqrt{1+\left( 1+\frac{2w_0^2}{(0.55C_n^2k^2 z)^{-6/5}}\right) \left(\frac{\lambda z}{\pi w_0^2} \right)^2
}$ is optical beamwidth at Rx, $w_0$ is optical beamwidth at Tx, $C_n^2$ is the index of refraction structure parameter, $\lambda$ is  the optical wave length, and $k$ is the optical wave number. 
Since the orientation fluctuations of UAV's is in the order of mrad, we can well approximate \eqref{df2} as
$r_d   \simeq \sqrt{r_{dx}^2 +    r_{dy}^2}$
where $r_{dx} = \theta_{tx}\,Z+x_{tx}+x_{rx}$ and $r_{dy}=\theta_{ty}\,Z+x_{ty}+x_{ry}$ which have Gaussian distribution as
$r_{dx}\sim\mathcal{N}(Z\theta_{tx}',\sigma_{rx}^2)$ and $r_{dy}\sim\mathcal{N}(Z\theta_{ty}',Z^2\sigma_{tyo}^2+\sigma_{typ}^2+\sigma_{ryp}^2)$ where $\sigma_{dx}^2=Z^2\sigma_{txo}^2+\sigma_{txp}^2+\sigma_{rxp}^2$ and 
$\sigma_{dy}^2=Z^2\sigma_{tyo}^2+\sigma_{typ}^2+\sigma_{ryp}^2$. 
From this,  the RV $r_d$ follows the Beckmann distribution \cite{beckmann1987scattering}
\begin{align}
\label{dc2}
&f_{r_d}(r_d) =  
\\
&~~~~~~\frac{r_d}{2\pi \sigma_{dx} \sigma_{dy}}\int_0^{2\pi}  e^{- \frac{(r_d\cos(\phi)-Z\theta_{tx}')^2}{2\sigma_{dx}^2} 
	-\frac{(r_d\sin(\phi)-Z\theta_{ty}')^2}{2\sigma_{dy}^2}
}\textrm{d}{\phi}. \nonumber
\end{align}
From \eqref{x3} and \eqref{dc2}, we obtain
\begin{align} 
\label{m1}   %
&f_{h_{pg}}(h_{pg}) =  \\ 
&\int_0^{2\pi}\frac{c_1 }{ h_{pg}} \exp\left(c_3\ln\left(\frac{A_0}{h_{pg}}\right)
+c_2\sqrt{\ln\left(\frac{A_0}{h_{pg}}\right)}\right) \textrm{d}\phi, \nonumber 
\end{align}
where $0<h_{pg}\leq A_0$, the parameters $c_1$, $c_2$, and $c_3$ are obtained in \eqref{xc} 
and $A_0=\frac{2r_a^2}{w_z^2} $.
Finally, from \eqref{a1}, \eqref{dx4}, \eqref{x5} and \eqref{m1}, and after some manipulations, the optical channel model under weak turbulence conditions, is derived in \eqref{xc2}.
%

%--------------------------------------
%--------------------------------------
\section{Prof of Theorem 3}
\label{Ape}
%--------------------------------------
%--------------------------------------
An approximation for Beckmann distribution is given in \cite{boluda2016novel}. From \cite{boluda2016novel}, we can approximate \eqref{dc2} as
\begin{align}
\label{kp1}
f_{r_d}(r_d) = \frac{r_d}{\sigma_m^2}\exp\left( - \frac{r_d^2}{2\sigma_m^2}  \right),
\end{align} 
where 
\begin{align}
\label{kp2}
\sigma_m^2 = \left( \frac{3Z^2\theta_{tx}'^2\sigma_{dx}^4      +         3Z^2\theta_{ty}'^2\sigma_{dy}^4      +     \sigma_{dx}^6      +       \sigma_{dy}^6}{2}  \right)^{\frac 13}.
\end{align}
From \eqref{x3} and \eqref{kp1}, we obtain
\begin{align} 
\label{kp3}   %
&f_{h_{pg}}(h_{pg}) =  \frac{ w_z^2  }{4\sigma_m^2 }   
A_0^{\frac{4\sigma_m^2}{w_z^2}}       h_{pg}^{\frac{w_z^2}{4\sigma_m^2}-1} .
~~~0\leq h_{pg}\leq A_0,
\end{align}
Finally, from \eqref{a1}, \eqref{x5} and \eqref{kp3}, and after some derivations, the closed-form channel model under weak turbulence condition is derived in \eqref{po1}.

%-----------------------------------------
%-----------------------------------------
%-----------------------------------------
\section{Prof of Theorem 4}
\label{Apee}
%-----------------------------------------
%-----------------------------------------
%-----------------------------------------
In some scenarios, UAVs have approximately same UAV's instability in the $x$ and $y$ axis. Under such conditions, we have
$\sigma_{txo}\simeq\sigma_{tyo}=\sigma_{to}$, $\sigma_{rxo}\simeq\sigma_{ryo}=\sigma_{ro}$, $\sigma_{txp}\simeq\sigma_{typ}=\sigma_{tp}$, and $\sigma_{rxp}\simeq\sigma_{ryp}=\sigma_{rp}$, and the  AoA of the received signal follows a Rician distribution as 
\begin{align}
\label{x2}
f_{\theta_a}(\theta_a) = \frac{\theta_a}{\sigma_{to}^2+\sigma_{ro}^2}
%\exp\left( -\frac{\theta_a^2+\theta_d^2}{2\left(\sigma_{to}^2+\sigma_{ro}^2 \right)} \right)
e^{-\frac{\theta_a^2+\theta_d^2}{2\left(\sigma_{to}^2+\sigma_{ro}^2 \right)}}
I_0\left(\frac{\theta_a\theta_d}{\sigma_{to}^2+\sigma_{ro}^2}\right),
\end{align}
where $I_0(.)$ is the modified Bessel function of the first kind with order zero, $\theta_a\in[0,\infty)$, and 
$\theta_d = \sqrt{\left(\theta_{tx}'+\theta_{rx}'\right)^2+\left(\theta_{ty}'+\theta_{ry}'\right)^2}$ is the boresight angle of received beam.
From \eqref{x1} and \eqref{x2}, we have
	%using \cite{wolframx}, we have
	\begin{align}
	\label{sfg1}
	&f_{h_{pa}}(h_{pa}) = \mathbb{M}\left(\frac{\theta_d}{\sqrt{\sigma_{to}^2+\sigma_{ro}^2}},
	\frac{\theta_\textrm{FOV}}{\sqrt{\sigma_{to}^2+\sigma_{ro}^2}}\right)\delta(h_{pa}) \\
	%-----------
	&~~~+\left(1  -   \mathbb{M}\left(\frac{\theta_d}{\sqrt{\sigma_{to}^2+\sigma_{ro}^2}},
	\frac{\theta_\textrm{FOV}}{\sqrt{\sigma_{to}^2+\sigma_{ro}^2}}\right)\right)\delta(h_{pa}-1), \nonumber
	\end{align}
	where  $\mathbb{M}(a,b)$ is the Marcum {\it Q}-function that is represented as \cite{wolframx}
	\begin{align}
	\label{op2}
	\mathbb{M}(a,b) = \int_b^\infty x \exp\left(-\frac{x^2+a^2}{2} \right) I_0(ax).
	\end{align}
Note that Marcum {\it Q}-function is a standard function that is available in popular mathematical software packages, e.g., MATLAB, and Mathematica.

	From \eqref{x3} and after some mathematical calculations, we obtain
	\begin{align}
	\label{x4}
	&f_{h_{pg}}(h_{pg}) = \tau \left(\frac{w_z^2}{2 r_a^2}\right)^{\tau}
	\exp{\left(-\frac{Z^2(\theta_{tx}'^2+\theta_{ty}'^2)}{2(Z^2\sigma_\textrm{to}^2+\sigma_\textrm{tp}^2 + \sigma_\textrm{rp}^2)}\right)} \nonumber \\
	%------------------------------
	&\times h_{pg}^{\tau-1} I_0\left(
	\sqrt{\frac{Z^2 w_z^2 (\theta_{tx}'^2+\theta_{ty}'^2)
			\ln\left(\frac{w_z^2}{2 r_a^2 h_{pg}}\right)}
		{2(Z^2\sigma_\textrm{to}^2+\sigma_\textrm{tp}^2 + \sigma_\textrm{rp}^2)^2}}\right) ,
	\end{align}
	where $h_{pg}\in\left[0,{2 r_a^2}/{w_z^2}\right]$ and $\tau=\frac{w_z^2}{4(Z^2\sigma_\textrm{to}^2+\sigma_\textrm{tp}^2 + \sigma_\textrm{rp}^2)}$. 
	Let us define $h'=h_l h_a h_{pg}$. The distribution of $h'$ is obtained as
	\begin{align}
	\label{x6}
	f_{h'}(h') = \int \frac{1}{h_lh_{a}} f_{h_{pg}}(h'/h_l h_a) f_{h_a}(h_a) d h_a.
	\end{align}
	Substituting \eqref{x5} and \eqref{x4} in \eqref{x6}, and after some simplifications, we obtain
	\begin{align}
	\label{x7}
	&f_{h'}(h') = \frac{\tau \exp(-r_o^2/2\sigma_d^2)}{ \sqrt{8\pi}\sigma_{L}}(h')^{-1}     
	\int_0^\infty         
	e^{-\tau x} \\
	&~~~~~~\times I_0\left(    \sqrt{\frac{r_o^2 w_z^2 x}{2\sigma_d^4}     }\right) 
	\exp\left( -\frac{\left(x-\ln \frac{\kappa h_l}{h'}-2\mu_{L}\right)^2}{8\sigma^2_{L}}\right) dx, \nonumber
	\end{align}
	where  
	$\kappa = \frac{2 r_a^2}{w_z^2}$, 
	$r_o=Z\sqrt{\theta_{tx}'^2+\theta_{ty}'^2}$, 
	$\sigma_d^2 = Z^2\sigma_\textrm{to}^2+\sigma_\textrm{tp}^2 + \sigma_\textrm{rp}^2$.
	Using the identity $I_\nu(z)=\sum_{k=0}^K\frac{1}{\Gamma(k+\nu+1)k!}\left(\frac{z}{2}\right)^{2k+\nu}$ \cite[eq. (03.02.02.0001.01)]{wolfra}, \eqref{x7} can be represented as
	\begin{align}
	\label{x8}
	&f_{h'}(h') = \frac{\tau e^{-r_o^2/2\sigma_d^2}}{ h' \sqrt{8\pi}\sigma_{L}}      
	\sum_{k=0}^K \frac{(r_o^2 w_z^2/8\sigma_d^4)^k}{\Gamma(k+1) k!} \int_0^\infty   x^k    \\  
	&~~~~~~~\times\exp\left( -\frac{\left(x-(\ln \frac{\kappa h_l}{h'}+2\mu_{L}) \right)^2     +    8\sigma^2_{L}\tau x}
	{8\sigma^2_{L}}\right) dx, \nonumber
	\end{align}
	where $\Gamma(.)$ is the Gamma function. In the following derivations we use the upper incomplete Gamma function $\Gamma(s,x)=\int_x^\infty t^{s-1}e^{-t}dt$ and lower incomplete Gamma function $\Upsilon(s,x)=\int_0^\infty t^{s-1}e^{-t}dt$ that are supported by MATLAB and Mathematica software packages. Using these and after some manipulations, when $h'\geq e^{q_2}$, the closed form expression for \eqref{x8} is obtained as 
	\begin{align}
	\label{x9}
	f_{h'}(h') =&~q_{3}   \frac{e^{ \tau\ln h'} }{h'}    
	\sum_{k=0}^K \sum_{j=0}^k q_1  
	\left({q_2  -\ln h'}\right)^{k-j} \\
	&\times \Gamma\left(\frac{j+1}{2},\frac{(q_2  -\ln h')^2}{8\sigma_L^2}\right), \nonumber
	\end{align}
	where the constant $q_1$, $q_2$, and $q_3$ are given in \eqref{z3}. Moreover, when $h'<e^{q_2}$, the closed form expression for \eqref{x8} is obtained as 
	\begin{align}
	\label{x11}
	&f_{h'}(h') = q_{3}   \frac{e^{ \tau\ln h'} }{h'}
	\sum_{k=0}^K \sum_{j=0}^k q_1  
	\left({q_2  -\ln h'}\right)^{k-j} \\
	&~~~~\times\left[\Gamma\left(\frac{j+1}{2},0\right) + (-1)^j\Upsilon\left(\frac{j+1}{2},\frac{(q_2  -\ln h')^2}{8\sigma_L^2}\right)\right] . \nonumber
	\end{align}
	Finally, using \eqref{sfg1}, \eqref{x9}, \eqref{x11}, and \cite[eq. (21)]{dabiri2019tractable}, the channel PDF is derived in \eqref{sfg}. 

%%%%%%%%%%%%%%%%%%%%%%%%%%% END PROOF %%%%%%%%%%%%%%%%%%%%%%%%%%%%%%%%%%%%%%%%%%%%%%%%%%%%%%%%%%%%%%%%%%%%%%%%%%%%%%%%%%%%%%%%%%%%%%%%%%%%%%%%%%%%%%%%%%%%%%%%%%%%

%------------------------------
%------------------------------
\section{Prof of Theorem 5}
\label{Apr}
%------------------------------
%------------------------------
For moderate to strong atmospheric turbulence conditions, $h_a$ can be well modeled by GG distribution as
\begin{eqnarray}
\label{fg1}
f_{\rm G}(h_a)=  \frac{2(\alpha\beta)^{\frac{\alpha+\beta}{2}}}{\Gamma(\alpha)\Gamma(\beta)} 
h_a^{\frac{\alpha+\beta}{2}-1}   
k_{\alpha-\beta}(2\sqrt{\alpha\beta h_a}), 
\end{eqnarray}
where $ \beta $ and $ \alpha $ are, respectively, the effective number of small-scale and large-scale eddies, which depend on Rytov variance $\sigma_R^2$, and $ k_{\nu}(.) $ is the modified Bessel function of the second kind of order $\nu$.
Based on \eqref{a1}, \eqref{dx4}, \eqref{m1} and \eqref{fg1}, and after some manipulations, the optical channel model under weak turbulence conditions, is derived in \eqref{cc1}.

%------------------------------------
%------------------------------------
\section{Prof of Theorem 6}
\label{Apre}
%------------------------------------
%------------------------------------
In the following derivation, we use the integral identity
\begin{align}
\label{c1}
&k_w(z) = \\ &\dfrac{\pi}{2\sin(\pi w)} \sum_{m=0}^M 
\left[\dfrac{(z/2)^{2m-w} }{\Gamma(m-w+1)m!}  -    \dfrac{(z/2)^{2m+w} }{\Gamma(m+w+1)m!} \right]. \nonumber
\end{align}
Based on \eqref{a1}, \eqref{dx4}, \eqref{m1}, substituting \eqref{c1} in \eqref{fg1}, using the results of \cite[Appendix C]{dabiri2019tractable}, and after some manipulations, the optical channel model under moderate to strong turbulence conditions, is derived in \eqref{b1}.

%------------------------------------
%------------------------------------
\section{Prof of Theorem 7}
\label{Apf}
%------------------------------------
%------------------------------------
Substituting \eqref{x4} and \eqref{fg1} in \eqref{x6}, using \eqref{c1}, applying a change of variable rule $y = \ln \frac{h'}{\kappa h_l h_a}$, and after some manipulations, we obtain
\begin{align}
\label{p1} 
&f_{h'}(h')=  B_0     \sum_{m=0}^M  h'^{m-1}   \\
%----------------------------------------------
&\times\Bigg[ g_{2m} h'^{\beta}   
\int_{\ln ({h'}/{\kappa h_l h_a})}^\infty
e^{(\tau-m-\beta)y} I_0\left(\frac{r_o}{\sigma_d^2}
\sqrt{\frac{-w_z^2 y}{2}}\right) dy \nonumber \\
%----------------------------------------------
&~-g_{3m}  h'^{\alpha}
\int_{\ln ({h'}/{\kappa h_l h_a})}^\infty
e^{(\tau-n-\alpha)y} I_0\left(\frac{r_o}{\sigma_d^2}
\sqrt{\frac{-w_z^2 y}{2}}\right) dy\Bigg], \nonumber
\end{align}
where $B_0 = \frac{\pi \tau \exp(-r_o^2/2\sigma_d^2)}{ \Gamma(\alpha)\Gamma(\beta)\sin(\pi (\alpha-\beta))}$.
In the following derivation, we utilize the identity \cite[eq. (2.32.2)]{jeffrey2007table}
\begin{align}
\label{p2}
\int e^{ax} x^k dx = e^{ax}\left(\sum_{j=0}^k\frac{(-1)^j j! \binom{k}{j}}{a^{j+1}} x^{k-j}\right).
\end{align}
Finally, using \eqref{sfg1}, \eqref{p1}, \eqref{p2}, \cite[eq. (03.02.02.0001.01)]{wolfra}, \cite[eq. (21)]{dabiri2019tractable}, and after some mathematical manipulations, the channel PDF is derived in \eqref{d1}.


\end{document}